\documentstyle[12pt, epsfig]{article}
\newcommand{\bea}   {\begin{eqnarray}}
\newcommand{\eea}   {\end{eqnarray}}
\begin{document}
\renewcommand{\thefootnote}{\fnsymbol{footnote}}

\thispagestyle{empty}
\title{On non-minimal ${\cal N}=4$ supermultiplets \\in $1D$ and their associated $\sigma$-models}
\author{Marcelo Gonzales\thanks{{\em e-mail: marcbino@cbpf.br}},
Sadi Khodaee\thanks{{\em e-mail: khodaee@cbpf.br}}
~and Francesco Toppan\thanks{{\em e-mail: toppan@cbpf.br}}
~\\~ \\
{\it CBPF, Rua Dr.}
{\it Xavier Sigaud 150,}
 \\ {\it cep 22290-180, Rio de Janeiro (RJ), Brazil.}\\}
\maketitle
\begin{abstract}
We construct the non-minimal linear representations of the ${\cal N}=4$ Extended Supersymmetry in one-dimension. They act on $8$ bosonic
and $8$ fermionic fields.
Inequivalent representations are specified by the mass-dimension of the fields and the connectivity of the associated graphs.
The oxidation to minimal ${\cal N}=5$ linear representations is given. Two types of ${\cal N}=4$ $\sigma$-models based on non-minimal
representations are obtained: the resulting off-shell actions are either manifestly invariant or depend on a constrained prepotential.
The connectivity properties of the graphs play a decisive role in discriminating inequivalent actions. These results find application
in partial breaking of supersymmetric theories.
\end{abstract}
\vfill
\rightline{CBPF-NF-001/10}

\newpage
\section{Introduction}

In this work we present a systematical investigation of the inequivalent non-minimal linear supermultiplets carrying a
representation
of the one-dimensional ${\cal N}=4$-Extended Superalgebra, as well as of their associated ${\cal N}=4$-invariant,
$1D$ $\sigma$-models.\par
The $1D$ ${\cal N}$-Extended Superalgebra, with ${\cal N}$ odd generators $Q_I$ ($I=1,2,\ldots, {\cal N}$)
and a single even generator $H$ satisfying the (anti)-commutation relations
\bea\label{sqm}
\{ Q_I, Q_J\}&=& \delta_{IJ} H, \nonumber\\
\relax [H, Q_I]&=&0,
\eea
is the superalgebra underlying the Supersymmetric Quantum Mechanics \cite{wit}.
In recent years the structure of its linear representations has been unveiled by a series of works (upon which
the present investigation is based) \cite{pt}--\cite{dfghilm2}, that we will briefly comment.\par
The linear representations under considerations (supermultiplets) contain a finite, equal number of bosonic and fermionic fields
depending on a single coordinate (the time). The operators $Q_I$ and $H$ act as differential operators. The
linear representations are characterized by a series of properties which, for sake of consistency, are reviewed in Appendix {\bf A}.
They include the {\em grading} of the fields (in physical terms, their {\em mass-dimension}), the {\em length} and
the {\em field content} of the supermultiplets \cite{{pt},{krt}}, the {\em dressing transformations}
\cite{{gr},{gr2},{pt}}, the association with graphs \cite{{fg},{dfghil},{dfghil2},{kt},{kt2},{top3}},
the {\em connectivity symbol} \cite{{kt},{kt2}} characterizing inequivalent representation with a given field content
(the notion of inequivalent representations has been discussed in \cite{{dfghil},{dfghil2}}),
the {\em mirror duality} \cite{top2}, etc.\par
The {\em minimal} linear representations (also called {\em irreducible supermultiplets}) are given by
the minimal number $n_{min}$ of bosonic (fermionic) fields for a given value of ${\cal N}$.
The value $n_{min}$ is given \cite{pt} by the formula
\begin{eqnarray}\label{irrepdim}
{\cal N}&=& 8l+m,\nonumber\\
n_{min}&=& 2^{4l}G(m),
\end{eqnarray}
where $l=0,1,2,\ldots$ and $m=1,2,3,4,5,6,7,8$.\par
$G(m)$ appearing in (\ref{irrepdim}) is the Radon-Hurwitz function
\begin{eqnarray}&\label{radonhur}
\begin{tabular}{|c|cccccccc|}\hline
$m $&$1$&$2$&$3$& $4$&$5$&$6$&$7$&$8$\\ \hline
$G(m)  $&$1$&$2$&$4$& $4$&$8$&$8$&$8$&$8$\\ \hline
\end{tabular}&\nonumber\\
&&\end{eqnarray}
Non-minimal linear representations have been discussed in \cite{{krt},{dfghilm},{dfghilm2},{gknty},{fkt}}.
The maximal {\em finite} number $n_{max}$ of bosonic (fermionic) fields entering a non-minimal representation
is given by \cite{{dfghilm},{dfghilm2}}
\bea
n_{max} &=& 2^{{\cal N}-1}.
\eea
An important subclass of non-minimal representations is given by the reducible but indecomposable supermultiplets
(see \cite{krt}). For this subclass the associated graph
(see Appendix {\bf A}) is connected (there is a path connecting any two given vertices).\par
For ${\cal N}=4$ we have that $n_{min}=4$ and $n_{max}=8$.  As a consequence, there are only two subclasses of non-minimal ${\cal N}=4$ representations. Besides the irreducible but indecomposable subclass, we have the subclass of fully reducible representations given
by the direct sum of two minimal ${\cal N}=4$ representations (the associated graph is disconnected and given
by two separate minimal
${\cal N}=4$ graphs).\par
In Section {\bf 2} we present the inequivalent non-minimal ${\cal N}=4$ supermultiplets, characterized
by their admissible field
contents and connectivity symbols (in subsection {\bf 2.1} we comment about the difference between the overall
connectedness of a graph and its associated connectivity symbol). We employ the techniques introduced in \cite{{kt},{kt2}}
and discussed in detail in those works. Therefore, here we limit ourselves to present the main results.\par
The investigation of the properties of the non-minimal ${\cal N}=4$ supermultiplets continues in Section {\bf 3},
where
we present the so-called ``oxidation diagrams'' connecting the non-minimal ${\cal N}=4$ representations with
the minimal ${\cal N}=5$ ones.\footnote{We postpone to the Conclusions the discussion about the meaning of
the term {\em oxidation} and of the physical importance of the so-called {\em oxidation program}, see \cite{{glpr},{top},{fkt}}.}
This Section employs the techniques introduced in \cite{kt2}.
The presented results answer the following question: which minimal ${\cal N}=5$ supermultiplets can be obtained by adding
an extra supersymmetry operator to a given non-minimal ${\cal N}=4$ supermultiplet in such a way to guarantee an overall
${\cal N}=5$ (\ref{sqm}) superalgebra.
We recall that, due to (\ref{irrepdim}), the minimal ${\cal N}=5$ representations contain the same number of fields
($8$ bosons and $8$ fermions) as the non-minimal ${\cal N}=4$
representations.\par
In Section {\bf 4} we present the most general unitary group commuting with the ${\cal N}=4$ supersymmetry operators
acting on a non-minimal supermultiplet. This Section applies the methods presented in \cite{fkt} where, for the minimal
supermultiplets, the supersymmetric extension of the Schur's lemma was constructed.\par
The next part of the paper deals with the construction of off-shell, ${\cal N}=4$-invariant supersymmetric $\sigma$-models
associated with each given non-minimal linear supermultiplet. $1D$ supersymmetric $\sigma$-models were first discussed
in \cite{{sigma1},{sigma2}}; minimal ${\cal N}=4$ $\sigma$-models were constructed in \cite{ikp}-\cite{three2},
while minimal ${\cal N}=8$-invariant $\sigma$-models were investigated in \cite{bikl}. There are also
supersymmetric $\sigma$-models based on non-linear realizations of the supersymmetry that we are not
discussing here (for a partial list of references one can consult \cite{fkt}).\par
The fields $x_j(t)$ ($j=1, \ldots, k$) of lower-dimension in a supermultiplet can be assumed \cite{krt}
to be bosonic and have $0$-mass dimension. They are physically interpreted \cite{{top},{grt},{fkt}} as the
target-coordinates of the associated ${\sigma}$-model. An ${\cal N}=4$-invariant off-shell action ${\cal S}$,
with the correct dimension of a kinetic term, is obtained \cite{{krt},{top},{grt},{fkt}} through
\bea\label{manifestn4}
{\cal S}&=& \int dt {\cal L}=\frac{1}{m}\int dt Q_1Q_2Q_3Q_4 F({\vec x})
\eea
(the dimensional parameter $m$ will be normalized to $1$ in the following),
where the supersymmetry operators $Q_i$'s act as graded derivatives and $F$ is the prepotential. By construction,
the action ${\cal S}$ is manifestly ${\cal N}=4$-invariant no matter which is the choice of $F$
(unconstrained prepotential). \par
In the case of a fully reducible supermultiplet given by the direct sum of two ${\cal N}=4$ irreducible
supermultiplets (whose $0$-mass dimension fields are denoted as ${\vec x}$, ${\vec y}$, respectively)
we have that interacting terms involving the fields belonging to the irreducible supermultiplets arise
provided that
\bea
F({\vec x}, {\vec y})&\neq& A({\vec x})+ B({\vec y}).
\eea
As a result, non-trivial interacting Lagrangians can be produced even from fully reducible representations
(which are trivial, from the representation theory point of view), therefore justifying the attention
we have to pay to them. \par
The (\ref{manifestn4}) manifest ${\cal N}=4$ construction has been discussed in \cite{grt}. In several cases,
however, this construction does not produce the most general ${\cal N}=4$
invariant action. The resulting Lagrangian can be of first order and furthermore, in the presence of fermionic
sources\footnote{A fermionic source \cite{{dfghil},{dfghil2},{kt}} is a fermionic $\frac{1}{2}$-mass dimension
field
which, in the graphical presentation of the supermultiplet, does not possess edges connecting it to the $0$-mass
dimension fields (the bosonic target coordinates). For ${\cal N}=4$, the connectivity symbol of a graph with $r$
fermionic sources is expressed as $r_4+\ldots$, see Appendix {\bf A}.},  not all fields belonging to the given
supermultiplet enter the Lagrangian.\par
On the other hand, even in those cases, the existence of a second-order Lagrangian involving all fields of the
supermultiplet is known \cite{{krt},{di}}. To systematically construct them a novel approach is here presented
(it will be referred as ``Construction II'', while (\ref{manifestn4}) will be referred as ``Construction I'').
Construction II is outlined as follows. For a reducible length-$3$ ${\cal N}=4$ representation of field content
$(k, 8,8-k)$, we consider at first (see the definitions in Appendix {\bf A}) its associated root supermultiplet
of length-$2$ (in a different context, the importance of invariant actions induced by the root supermultiplets
has also been discussed in \cite{{gauging1},{gauging2},{gauging3}}). The root supermultiplet contains $8$ bosonic
fields $x_i$ (the target coordinates) and $8$ fermionic fields $\psi_i$. A Lagrangian ${\cal L}$ for the root
supermultiplet is at first constructed by setting, as in (\ref{manifestn4}),
\bea\label{Wprep}
{\cal L}&=& Q_1Q_2Q_3Q_4 W({\vec x}),
\eea
where the prepotential $W({\vec x})$ is now function of the $8$ bosonic fields.\par
An equivalent, up to a total derivative, Lagrangian ${\overline {\cal L}}$ functionally depends on the fields
and their first-order
time-derivatives alone. Therefore
\bea
{\overline {\cal L}}&\equiv& {\overline {\cal L}}(x_i, {\dot x_i},\psi_i,{\dot\psi_i}).
\eea
The next step consists in constraining ${\overline{\cal L}}$ such that, for $j=k+1,\ldots, 8$, we have
\bea\label{dressconstr}
\frac{\partial{\overline {\cal L}}}{\partial x_j}&=& 0,
\eea
eliminating its dependence on $x_j$'s. This condition allows us to regard, according to the dressing procedure,
the ${\dot x_j}$'s no longer as derivative fields, but as the auxiliary fields $g_j$ of mass-dimension $1$
entering the $(k,8,8-k)$ supermultiplet. We can therefore set
\bea\label{renaming}
g_j&=&{\dot x_j}\nonumber\\
{\overline {\cal L}}&\equiv& {\overline {\cal L}} (x_l, {\dot x_l},\psi_i, {\dot \psi_i}, g_j),
\eea
($l=1,\ldots, k$, while $i=1,\ldots, 8$ and $j=k+1,\ldots, 8$).\par
Setting (\ref{renaming}) is not something innocuous. One has in fact to guarantee that the resulting action,
after the ``renaming'' of the fields,
is still ${\cal N}=4$-invariant. Together with (\ref{dressconstr}), this requirement produces a constraint on
the prepotential $W(x_l)$. In the following we will compare the invariant actions arising from the
constructions I and II and discuss the constraints on the prepotential.\par
To derive the off-shell invariant actions we implemented a special package for {\em Maple 11}.
For convenience we had to use different (but equivalent) conventions for the presentations of the non-minimal
supermultiplets with respect to the explicit construction given in Appendix {\bf B}.
We present in Section {\bf 5} and {\bf 6} some selected cases which exemplify the general picture.
In Section {\bf 5} we discuss the
construction I. In Section {\bf 6} we discuss the construction II. It will be manifest that inequivalent
${\cal N}=4$-invariant actions are obtained for supermultiplets
presenting the same field content, but differing in connectivity symbol. In the Conclusions we make comments
on the obtained
results.\par
The paper is complemented,  for sake of clarity, with the presentation of a few selected (unoriented, color-blind)
graphs associated to non-minimal supermultiplets.

\section{Non-minimal ${\cal N}=4$ supermultiplets}

We present here the complete list of inequivalent ${\cal N}=4$ non-minimal supermultiplets associated to
connected graphs (therefore providing
reducible but indecomposable representations). \par
Up to equivalence, there exists a unique length-$2$ ${\cal{N}}=4$ root supermultiplet, based on a connected
graph, of field content $(8,8)$ (it is explicitly given in Appendix {\bf B}).\par
The inequivalent non-minimal supermultiplets of length-$3$ are given by the table below. One should note
that, for field content $(k,8,8-k)$ with
$k=2,3,4,5,6$, inequivalent supermultiplets are discriminated by their respective connectivity symbol
(see Appendix {\bf A}) and are named according to the given label. The table further reports the dually
related supermultiplet (see Appendix {\bf A}). We have
\begin{center}
\begin{tabular}{|c|c|c|c|}\hline
\emph{field content:}& \emph{label:}& \emph{connectivity symbol:}&\emph{dual supermultiplet:}\\ \hline
$(1,8,7)$ && $4_4 + 4_3$ & $(7,8,1)$ \\ \hline
$(2, 8, 6)$& $A$& $2_4 + 4_3 + 2_2$& $(6,8,2)_A$ \\ \cline{2-4}
& $B$& $8_3$ & $(6,8,2)_B$ \\ \hline
$(3, 8, 5)$ &$A$& $1_4 +3_3 + 3_2 + 1_1$& $(5,8,3)_A$ \\ \cline {2-4}
&$B$& $4_3 + 4_2$& $(5,8,3)_B$ \\ \hline
$(4, 8, 4)$& $A$& $1_4 + 6_2 + 1_0$ & self-dual \\ \cline {2-4}
&$B$& $4_3 + 4_1$ & self-dual \\ \cline {2-4}
&$C$& $2_3 + 4_2 +2_1$ & self-dual \\ \cline {2-4}
&$D$& $8_2$ & self-dual \\ \hline
$(5, 8, 3)$& $A$& $1_3 +3_2 + 3_1 + 1_0$& $(3,8,5)_A$ \\ \cline{2-4}
& $B$& $4_2 + 4_1$ & $(3,8,5)_B$ \\ \hline
$(6, 8, 2)$& $A$& $2_2 + 4_1 + 2_0$ & $(2,8,6)_A$ \\ \cline{2-4}
& $B$& $8_1$ & $(2,8,6)_B$ \\ \hline
$(7, 8, 1)$ && $4_1 + 4_0$& $(1,8,7)$  \\ \hline
\end{tabular}
\end{center}
\bea\label{red}&&\eea
The explicit construction of a representative supermultiplet in each given inequivalent class is given
in the Appendix {\bf B}.\par
We further complete the list by presenting the non-minimal (reducible but indecomposable) ${\cal N}=4$
representations with length $l>3$. They are uniquely characterized by their field content. They
are given by
\bea
&
(1,4,6, 4, 1),\quad (1, 7, 7,1),\quad (1,4, 7, 4)\leftrightarrow (4, 7, 4, 1),\quad
(1 ,5, 7, 3)\leftrightarrow (3, 7, 5, 1),
& \nonumber\\
& (1, 6, 7, 2)\leftrightarrow (2, 7, 6, 1),\quad (2, 6, 6, 2),&
\eea
with dually related supermultiplets connected by arrows.


\subsection{Connectivity symbol of the ${\cal N}=4$ fully reducible representations}

It is interesting to compare the connectivity symbol of the non-minimal (reducible,
but indecomposable) linear representations of the one-dimensional ${\cal N}=4$ Extended Superalgebra,
with the connectivity symbol of the fully reducible ${\cal N}=4$ representations involving $8$ bosonic
and $8$ fermionic fields. The representations are otained by direct sum of two $(k,4,4-k)$ (for $k=0,1,2,3,4$)
${\cal N}=4$
irreducible (minimal) representations. \par
For length-$3$ fully reducible representations we have the table\footnote{In order to distinguish the
fully reducible
representations of field content $(8,8)\equiv (4,4,0)\oplus(0,4,4)$, $(7,8,1)$, $(1,8,7)$ from their
non-minimal, reducible but indecomposable counterparts, a label is introduced (respectively, either ``$FR$'' or ``$red$'').
The fully reducible representations will be denoted as $(8,8)_{FR}$, $(7,8,1)_{FR}$, $(1,8,7)_{FR}$; the reducible but
indecomposable supermultiplets entering table (\ref{red}) will be denoted as
$(8,8)_{red}$, $(7,8,1)_{red}$, $(1,8,7)_{red}$.}
\begin{center}
\begin{tabular}{|c|c|c|c|}\hline
\emph{field content:}& \emph{label:}& \emph{decomposition:}&\emph{connectivity symbol:}\\ \hline
$(1,8,7)$ &$FR$&  $(1,4,3)\oplus (0,4,4)$ &$4_4 + 4_3$  \\ \hline
$(2, 8, 6)$& $a$&$(2,4,2)\oplus(0,4,4)$& $4_4 + 4_2$ \\
& $b$& $(1,4,3)\oplus(1,4,3)$&$8_3$  \\ \hline
$(3, 8, 5)$ &$a$&$(3,4,1)\oplus(0,4,4)$ &$4_4 +4_1$ \\
&$b$&$(2,4,2)\oplus(1,4,3)$& $4_3 + 4_2$\\ \hline
$(4, 8, 4)$& $a$& $(4,4,0)\oplus(0,4,4)$&$4_4 + 4_0$  \\
&$b$& $(3,4,1)\oplus(1,4,3)$&$4_3+4_1$  \\
&$c$&$(2,4,2)\oplus(2,4,2)$ &$8_2$  \\
 \hline
$(5, 8, 3)$& $a$& $(4,4,0)\oplus(1,4,3)$&$4_4+4_3$  \\
& $b$& $(3,4,1)\oplus(2,4,2)$&$4_2 + 4_1$  \\ \hline
$(6, 8, 2)$& $a$&$(4,4,0)\oplus(2,4,2)$& $4_2 + 4_0$ \\
& $b$& $(3,4,1)\oplus(3,4,1)$&$8_1$  \\ \hline
$(7, 8, 1)$ &$FR$& $(4,4,0)\oplus(3,4,1)$&$4_1 + 4_0$  \\ \hline
\end{tabular}
\end{center}
\bea\label{FRmult}&&\eea
A comment is in order. The reducible but indecomposable representations discussed before
and the fully reducible representations here listed are inequivalent. Indeed, their associated graphs are
connected
in the first case, disconnected in the second case. The so-called {\em connectivity symbol} introduced in \cite{kt}
allows to discriminate the inequivalent representations inside each broad class of fully reducible
or reducible but indecomposable representations. On the other hand, in several cases the connectivity symbol {\em alone}
cannot discriminate between a fully reducible and its reducible but indecomposable counterpart.
For instance, the fully reducible representation $(2,8,6)_b$ possesses the same connectivity symbol as the reducible
but indecomposable representation $(2,8,6)_B$. Once more, the extra information allowing to discriminate the two is
the overall connectedness of the associated graph (either connected or disconnected).

\section{The $({\cal N}=4)\Rightarrow ({\cal N}=5)$ oxidation}

Non-minimal ${\cal N}=4$ representations (both the reducible but indecomposable and the fully reducible ones
acting on $8$ bosonic and $8$ fermionic fields) can be ``oxidized'' to ${\cal N}=5$ minimal representations,
adding a $5^{th}$ supersymmetry transformation compatible with the $4$ previous supersymmetry transformations.
\par
The minimal ${\cal N}=5$ representations have been classified in \cite{kt}. There are inequivalent length-$3$
${\cal N}=5$ representations of field content $(k,8,8-k)$, for $k=2,3,4,5,6$, which are discriminated by their
connectivity symbol. \par
We present a series of tables specifying which minimal ${\cal N}=5$ representation (in a column) results from
the oxidation of a non-minimal ${\cal N}=4$ representation (in a row). A positive answer is marked by an ``$X$''.
The representations are expressed in terms of their connectivity symbol. The ${\cal N}=4$ reducible but indecomposable
representations (associated with connected graphs) appear in the upper rows; the ${\cal N}=4$ fully reducible
representations (associated with disconnected graphs) appear in the lower rows.
\par
We get the following oxidation tables, for each given field content $(k,8,8-k)$ with $k=2,3,4,5,6$.\par

\par

For $(2,8,6)$ we have
\begin{eqnarray}
\begin{array}{c|c|c|c|}
\cline{2-4}
&     ({{\cal N}=4})\Rightarrow ({{\cal N}=5)}:   &   2_5+2_4+4_3   &   6_4+2_3   \\
\cline{1-4}
\multicolumn{1}{|c}{}{{\em Connected:}\quad} \vline&              2_4+4_3+2_2                        &       X         &      X  \\
\cline{2-4}
\multicolumn{1}{|c}{}                             \vline&                8_3                              &                 &      X      \\
\cline{1-4}
\multicolumn{1}{|c}{}
{{\em Disconnected:}}    \vline&              4_4+4_2                            &       X         &             \\
\cline{2-4}
\multicolumn{1}{|c}{}                            \vline&                8_3                              &                 &      X      \\
\hline
\end{array}
\end{eqnarray}

For $(3,8,5)$ we have
\begin{eqnarray}
\begin{array}{c|c|c|c|}
\cline{2-4}
                                                   &    ({{\cal N}=4})\Rightarrow ({{\cal N}=5)}:   &  1_5+3_4+4_2  &    2_4+5_3+1_2\\
\cline{1-4}
\multicolumn{1}{|c}{}{{\em Connected:\quad}} \vline&            1_4+3_3+3_2 +1_1 &     X     & X  \\
\cline{2-4}
\multicolumn{1}{|c}{}        \vline&                                4_3+4_2 &      &          X   \\
\cline{1-4}
\multicolumn{1}{|c}{}{\em Disconnected:}   \vline&   4_4+4_1       &  X         &   \\
\cline{2-4}
\multicolumn{1}{|c}{}                            \vline&             4_3+4_2  &           &             X  \\
\hline
\end{array}
\end{eqnarray}

For $(4,8,4)$ we have
\begin{eqnarray}
\begin{array}{c|c|c|c|c|}
\cline{2-5}
                                &           ({{\cal N}=4})\Rightarrow ({{\cal N}=5)}:&4_4+4_1 &1_4+3_3+3_2+1_1&4_3+4_2\\
\cline{1-5}
\multicolumn{1}{|c}{}{{\em Connected:\quad}}  \vline&     1_4+6_2+1_0    &      &  X    &  \\
\cline{2-5}
\multicolumn{1}{|c}{}  \vline&4_3+4_1&X&& X \\
\cline{2-5}
\multicolumn{1}{|c}{}      \vline&2_3+4_2+2_1  &   &   X  &  X \\
\cline{2-5}
 \multicolumn{1}{|c}{}        \vline&  8_2  &  &  &  X \\
\cline{1-5}
\multicolumn{1}{|c}{}{{\em Disconnected:}}      \vline&           4_4+4_0 &   X       &   &  \\
\cline{2-5}
\multicolumn{1}{|c}{}  \vline&  4_3+4_1&&X&  \\
\cline{2-5}
\multicolumn{1}{|c}{}  \vline&   8_2&&&X \\
\hline
\end{array}
\end{eqnarray}

For $(5,8,3)$ we have
\begin{eqnarray}
\begin{array}{c|c|c|c|}
\cline{2-4}
                             &              ({{\cal N}=4})\Rightarrow ({{\cal N}=5)}:   &   4_3+3_1+1_0  & 1_3+5_2+2_1    \\
\cline{1-4}
\multicolumn{1}{|c}{}{\em Connected:\quad} \vline& 1_5+3_2+3_1+1_0     &   X   &   X  \\
\cline{2-4}
\multicolumn{1}{|c}{}   \vline &       4_2+4_1     &   &       X  \\
\cline{1-4}
\multicolumn{1}{|c}{}{\em Disconnected:}       \vline&       4_3+4_0     &   X   &       \\
\cline{2-4}
\multicolumn{1}{|c}{}       \vline&           4_2+4_1 &       &   X  \\
\hline
\end{array}
\end{eqnarray}

For $(6,8,2)$ we have
\begin{eqnarray}
\begin{array}{c|c|c|c|}
\cline{2-4}
                &    ({{\cal N}=4})\Rightarrow ({{\cal N}=5)}:   &4_2+2_1+2_0   &2_2+6_1\\
\cline{1-4}
\multicolumn{1}{|c}{}{\em Connected:\quad}  \vline&    2_2+4_1+2_0  &  X  & X  \\
\cline{2-4}
 \multicolumn{1}{|c}{}   \vline&   8_1 &   &  X  \\
\cline{1-4}
\multicolumn{1}{|c}{}{\em Disconnected:}  \vline&         4_2+4_0  &     X        &    \\
\cline{2-4}
\multicolumn{1}{|c}{}              \vline&     8_1 &   &    X  \\
\hline
\end{array}
\end{eqnarray}
\par

Both the reducible but indecomposable and the fully reducible non-minimal ${\cal N}=4$ representations of field content
$(8,8)$, $(1,8,7)$ and $(7,8,1)$ are oxidized to the minimal ${\cal N}=5$ representations which are uniquely specified by
the corresponding
field content $(8,8)$, $(1,8,7)$ and $(7,8,1)$.\par
Combining these results with the results presented in \cite{kt2} we get that all non-minimal length-$3$ representations are
oxidized to the maximal number ${\cal N}_{max}=8$ of extended supersymmetries.\par
The maximal number ${\cal N}_{max}$ of supersymmetries operators (oxidized supersymmetries) acting on non-minimal
${\cal N}=4$
representations of length $l=4,5$ is given by the table
\begin{center}
\begin{tabular}{|c|c|}\hline
\emph{field content:}&${\cal N}_{max}$\emph{:}\\ \hline
$(1, 7, 7,1)$ & $7$ \\ \hline
$(2, 6, 6, 2)$ & $6$ \\ \hline
$(1, 6, 7, 2)\leftrightarrow (2, 7, 6, 1)$ & $6$ \\ \cline{1-1}\hline
$(1 ,5, 7, 3)\leftrightarrow (3, 7, 5, 1)$ & $5$ \\ \cline{1-1}\hline
$(1,4, 7, 4)\leftrightarrow (4, 7, 4, 1)$ & $4$ \\ \cline{1-1}\hline
$(1,4,6, 4, 1)$ & $4$ \\ \hline
\end{tabular}
\end{center}
\bea&&\eea

\section{Non-minimal ${\cal N}=4$ supermultiplets and invariant groups}

The Schur's lemma states that the irreducible representations of the Clifford algebras are of three types
(real, almost complex or quaternionic), according to the most general matrix commuting with all Clifford generators
(see \cite{oku}). Since minimal root supermultiplets (see \cite{pt}) are uniquely determined by their associated Euclidean
Clifford algebra, they inherit the corresponding Schur's property. The determination of the Schur's property of
higher-length supermultiplets requires the compatibility with the dressing (this implies that the most general
commuting matrix of the root supermultiplet is restricted by the further requirement of commuting with the dressing
matrix $D$ discussed in Appendix {\bf A}).
The analysis for the minimal supermultiplets has been presented in \cite{fkt}. We extend here the investigation of
\cite{fkt} to the case of the non-minimal ${\cal N}=4$ supermultiplets.\par
We are looking for the most general real-valued antisymmetric matrix $\Sigma$ of the form
$
\Sigma =\left(
\begin{array}{cc}
\Sigma^{up}&0\\
0&\Sigma^{down}
\end{array}
\right)
$, where $\Sigma^{up}$ ($\Sigma^{down}$) is an $8\times 8$ matrix acting on the bosonic (respectively, fermionic)
fields, commuting with the $4$ non-minimal supersymmetry operators
${\widehat Q}_I$, i.e.
\bea
[{\widehat Q}_I, \Sigma]&=&0, \quad for\quad I=1,2,3,4.
\eea
For the non-minimal, reducible but indecomposable $(8,8)$ root supermultiplet, the most general matrix $\Sigma$
can be written as $\Sigma=\sum_{i=1}^{i=6}\lambda_i{\Sigma_i}$, where
the matrices ${\Sigma_i}={\Sigma_i}^{up}\oplus {\Sigma_i}^{down}$ are the generators of the $su(2)\oplus su(2)$
Lie algebra.\par
 Without loss of generality we can work with the conventions of Appendix {\bf B}. We have, explicitly,
 \bea&
\begin{array}{lll}
{\Sigma_1}^{up}= \tau_2\otimes{\tau}_2\otimes\tau_A,
&
{\Sigma_2}^{up}= {\bf 1}_2\otimes{\tau_A }\otimes{\bf 1}_2,
& {\Sigma_3}^{up}= \tau_2\otimes \tau_1\otimes{\tau}_A,\\
 {\Sigma_4}^{up}= {\tau}_A\otimes\tau_1\otimes{\bf 1}_2,
&{\Sigma_5}^{up}= {\tau}_1\otimes{\bf 1}_2\otimes\tau_A,
&{\Sigma_6}^{up}= \tau_A\otimes\tau_2\otimes{\bf 1}_2,
\end{array}
&
\eea
while ${\Sigma_i}^{down}= {\Sigma_i}^{up}$ for $i=1,2,3$ and ${\Sigma_i}^{down}= -{\Sigma_i}^{up}$ for $i=4,5,6$.
\par
The unitary invariant groups, commuting with the $4$ supersymmetry operators of the length-$3$, reducible
but indecomposable, non-minimal supermultiplets are given by the table
\bea
&\begin{array}{|c|c|}\hline
\emph{supermultiplet:} & \emph{commuting group:}\\ \hline
(2, 8, 6)_A &   U(1)\\ \hline
(2, 8, 6)_B &  {\bf 1}\\ \hline
(4, 8, 4)_A & {\bf 1}\\ \hline
(4, 8, 4)_B & SU(2)\\ \hline
(4, 8, 4)_C &  {\bf 1}\\ \hline
(4, 8, 4)_D &  U(1)\otimes U(1)\\ \hline
(6, 8, 2)_A & U(1)\\ \hline
(6, 8, 2)_B &  {\bf 1}\\ \hline
\end{array}&
\eea
In the remaining cases, for field content $(k,8,8-k)$ with $k$ odd, the most general unitary group is just the
identity group ${\bf 1}$.\par
A similar table can be produced for length $l=2,3$ ${\cal N}=4$ non-minimal, fully reducible supermultiplets.
We have
\bea
&\begin{array}{|l|c|}\hline
\emph{supermultiplet:} & \emph{commuting group:}\\ \hline
(8, 8)_{FR} &   SU(2)\otimes SU(2)\\ \hline
(1,8, 7)_{FR} &   SU(2)\otimes {\bf 1}_2\\ \hline
(2, 8, 6)_a &   SU(2)\otimes U(1)\\ \hline
(2, 8, 6)_b &  {\bf 1}_2\otimes {\bf 1}_2\\ \hline
(3, 8,5)_{a} &   SU(2)\otimes {\bf 1}_2\\ \hline
(3, 8,5)_{b} &   U(1)\otimes {\bf 1}_2\\ \hline
(4, 8, 4)_a & SU(2)\otimes SU(2)\\ \hline
(4, 8, 4)_b & {\bf 1}_2\otimes {\bf 1}_2\\ \hline
(4, 8, 4)_c &  U(1)\otimes U(1)\\ \hline
(5, 8,3)_{a} &   SU(2)\otimes {\bf 1}_2\\ \hline
(5, 8,3)_{b} &   U(1)\otimes{\bf 1}_2\\ \hline
(6, 8, 2)_a & SU(2)\otimes U(1)\\ \hline
(6, 8, 2)_b &  {\bf 1}_2\otimes{\bf 1}_2\\ \hline
(7,8, 1)_{FR} &   SU(2)\otimes {\bf 1}_2\\ \hline
\end{array}&
\eea

\section{Manifestly ${\cal N}=4$ $\sigma$-models for non-minimal supermultiplets}

For the following length-$3$, non-minimal, reducible but indecomposable supermultiplets, the Construction I
(see (\ref{manifestn4})) of the ${\cal N}=4$ off-shell invariant actions produces a first-order Lagrangian:
\bea\label{firstordermult}
&(1,8,7)_{red}, (2,8,6)_A, (3,8,5)_A, (4,8,4)_A, (4,8,4)_B.&
\eea
With the only exception of $(4,8,4)_B$, these are the supermultiplets admitting, see the footnote in the Introduction,
fermionic sources. \par
In the remaining cases, namely for
\bea\label{secondordermult}
&(2,8,6)_B, (3,8,5)_B, (4,8,4)_C, (4,8,4)_D, (5,8,3)_A,\nonumber\\
&(5,8,3)_B, (6,8,2)_A, (6,8,2)_B, (7,8,1)_{red},
\eea
 the Construction I produces second-order Lagrangians.
\par
For each supermultiplet entering (\ref{secondordermult}), the Constructions I and II (detailed in the Introduction) produce,
up to a
total derivative, the same Lagrangian. Therefore, the corresponding actions are (as a consequence of Construction I)
manifestly ${\cal N}=4$-invariant and depend on an unconstrained prepotential. We present, for a few selected cases,
the explicit computation of the Lagrangian. We write down the invariant Lagrangian, up to a total derivative,
for $(4,8,4)_C$
(with connectivity symbol $2_3+4_2+2_1$)
and $(2,8,6)_B$. We compare the latter result with the Lagrangian
obtained from Construction I applied to the fully reducible supermultiplet $(2,8,6)_b$ (see (\ref{FRmult})),
characterized by the same connectivity symbol, $8_3$, as $(2,8,6)_B$.\par
The component fields (the $\upsilon$'s, barred or otherwise, denote the target coordinates, the $\lambda$'s
the fermionic fields and the $g$'s the auxiliary fields) are respectively given by
$(\upsilon_{1},\upsilon_{2},\upsilon_{3},
\bar{\upsilon}_{1};\lambda_{0},\lambda_{i},\bar{\lambda}_{0},\bar{\lambda}_{i};g_{0},\bar{g}_{0},\bar{g}_{2},\bar{g}_{3})$,
with $i=1,2,3$, for $(4,8,4)_C$ and by\\ $(\upsilon_{0},\bar{\upsilon}_{0};\lambda_{0},\lambda_{i},\bar{\lambda}_{0},\bar{\lambda}_{i};g_{i},\bar{g}_{i})$,
with $i=1,2,3$,
for both $(2,8,6)_B$ and $(2,8,6)_b$.\\
The supersymmetry transformations are here explicitly obtained by dressing the
${\cal N}=8$ root supermultiplet
expressed
in terms of the octonionic structure constants $C_{ijk}$, see \cite{{crt},{krt}}
\bea
\widehat{Q}_{i}(\upsilon_{0},\upsilon_{j};\lambda_{0},\lambda_{j})&=&(\lambda_{i},-\delta_{ij}\lambda_{0}-
C_{ijk}\lambda_{k};-\dot{\upsilon}_{i},\delta_{ij}\dot{\upsilon}_{0}+C_{ijk}\dot{\upsilon}_{k}),\nonumber\\
\widehat{Q}_{8}(\upsilon_{0},\upsilon_{j};\lambda_{0},\lambda_{j})&=&(\lambda_{0},\lambda_{j};
\dot{\upsilon}_{0},\dot{\upsilon}_{j}),
\eea
where $i=1,..,7$ and the fields have been renamed in such a way to respect the ${\cal N}=4$ quaternionic subalgebra.
For $(2,8,6)_B$  and $(2,8,6)_b$ the associated graphs are given in, respectively, figure $4$ and $6$. \\
We have the following results (the Einstein convention over repeated indexes is understood):\par
For $(4,8,4)_C$ the Lagrangian is
\begin{eqnarray}\label{484C}
{\cal L}&=&\Phi(\dot{\upsilon}_{1}^{2}+
\dot{\upsilon}_{2}^{2}+\dot{\upsilon}_{3}^{2}+
\dot{\bar{\upsilon}}_{1}^{2}+g_{0}^{2}+
\bar{g}_{0}^{2}+\bar{g}_{2}^{2}+\bar{g}_{3}^{2}+
\dot{\lambda}_{0}\lambda_{0}+\dot{\bar{\lambda}}_{0}\bar{\lambda}_{0}+\\\nonumber
&&\dot{\lambda}_{1}\lambda_{1}+
\dot{\lambda}_{2}\lambda_{2}+\dot{\lambda}_{3}\lambda_{3}+
\dot{\bar{\lambda}}_{1}\bar{\lambda}_{1}+
\dot{\bar{\lambda}}_{2}\bar{\lambda}_{2}+\dot{\bar{\lambda}}_{3}\bar{\lambda}_{3})+\\\nonumber
&&\Phi_{1}[\dot{\upsilon}_{3}
(\bar{\lambda}_{0}\bar{\lambda}_{2}+
\lambda_{2}\lambda_{0}+\bar{\lambda}_{1}\bar{\lambda}_{3}+
\lambda_{1}\lambda_{3})-\dot{\upsilon}_{2}(\bar{\lambda}_{0}
\bar{\lambda}_{3}+\lambda_{3}\lambda_{0}-\bar{\lambda}_{1}\bar{\lambda}_{2}-\lambda_{1}\lambda_{2})+\\\nonumber
&&+g_{0}(\bar{\lambda}_{2}\bar{\lambda}_{3}-
\lambda_{2}\lambda_{3}-\bar{\lambda}_{0}
\bar{\lambda}_{1}+\lambda_{1}\lambda_{0})+
\bar{g}_{0}(\lambda_{0}\bar{\lambda}_{1}+
\lambda_{1}\bar{\lambda}_{0}+\lambda_{2}\bar{\lambda}_{3}-\lambda_{3}\bar{\lambda}_{2})+\\\nonumber
&&\bar{g}_{2}(\lambda_{2}\bar{\lambda}_{1}+
\lambda_{1}\bar{\lambda}_{2}-\lambda_{0}
\bar{\lambda}_{3}+\lambda_{3}\bar{\lambda}_{0})+
\bar{g}_{3}(\lambda_{0}\bar{\lambda}_{2}-\lambda_{2}\bar{\lambda}_{0}+
\lambda_{3}\bar{\lambda}_{1}+\lambda_{1}\bar{\lambda}_{3})]+\\\nonumber
&&\Phi_{2}[\dot{\upsilon}_{3}(\bar{\lambda}_{0}
\bar{\lambda}_{1}+\lambda_{1}\lambda_{0}+
\bar{\lambda}_{2}\bar{\lambda}_{3}+\lambda_{2}\lambda_{3})+
\dot{\upsilon}_{1}(\bar{\lambda}_{0}\bar{\lambda}_{3}+
\lambda_{3}\lambda_{0}-\bar{\lambda}_{1}\bar{\lambda}_{2}-\lambda_{1}\lambda_{2})+\\\nonumber
&&+g_{0}(\bar{\lambda}_{3}\bar{\lambda}_{1}-\lambda_{3}\lambda_{1}-
\bar{\lambda}_{0}\bar{\lambda}_{2}+\lambda_{2}\lambda_{0})+
\bar{g}_{0}(\lambda_{0}\bar{\lambda}_{2}+\lambda_{2}\bar{\lambda}_{0}+
\lambda_{2}\bar{\lambda}_{3}-\lambda_{3}\bar{\lambda}_{2})+\\\nonumber
&&-\bar{g}_{2}(\lambda_{0}\bar{\lambda}_{0}+
\lambda_{1}\bar{\lambda}_{1}-\lambda_{2}\bar{\lambda}_{2}+
\lambda_{3}\bar{\lambda}_{3})+\bar{g}_{3}(\lambda_{3}
\bar{\lambda}_{2}+\lambda_{2}\bar{\lambda}_{3}-
\lambda_{0}\bar{\lambda}_{1}+\lambda_{1}\bar{\lambda}_{0})] + \\\nonumber
&&\Phi_{3}[\dot{\upsilon}_{2}
(\bar{\lambda}_{0}\bar{\lambda}_{1}+\lambda_{1}\lambda_{0}-
\bar{\lambda}_{2}\bar{\lambda}_{3}-\lambda_{2}\lambda_{3})-
\dot{\upsilon}_{1}(\bar{\lambda}_{0}\bar{\lambda}_{2}+\lambda_{2}\lambda_{0}+
\bar{\lambda}_{1}\bar{\lambda}_{3}+\lambda_{1}\lambda_{3})+\\\nonumber
&&+g_{0}(\bar{\lambda}_{1}\bar{\lambda}_{2}-\lambda_{1}\lambda_{2}-
\bar{\lambda}_{0}\bar{\lambda}_{3}+\lambda_{3}\lambda_{0})+
\bar{g}_{0}(\lambda_{0}\bar{\lambda}_{3}+\lambda_{3}
\bar{\lambda}_{0}+\lambda_{1}\bar{\lambda}_{2}-\lambda_{2}\bar{\lambda}_{1})+\\\nonumber
&&\bar{g}_{2}(\lambda_{0}\bar{\lambda}_{1}-
\lambda_{1}\bar{\lambda}_{0}+\lambda_{2}\bar{\lambda}_{3}+
\lambda_{3}\bar{\lambda}_{2})-\bar{g}_{3}(\lambda_{0}
\bar{\lambda}_{0}+\lambda_{1}\bar{\lambda}_{1}+\lambda_{2}\bar{\lambda}_{2}-\lambda_{3}\bar{\lambda}_{3})] + \\\nonumber
&&\Phi_{\bar{1}}[\dot{\upsilon}_{1}
(\lambda_{0}\bar{\lambda}_{0}-\lambda_{1}\bar{\lambda}_{1}+
\lambda_{2}\bar{\lambda}_{2}+\lambda_{3}\bar{\lambda}_{3})-
\dot{\upsilon}_{2}(\lambda_{0}\bar{\lambda}_{3}-
\lambda_{3}\bar{\lambda}_{0}+\lambda_{2}\bar{\lambda}_{1}+\lambda_{1}\bar{\lambda}_{2})+\\\nonumber
&&\dot{\upsilon}_{3}(\lambda_{0}\bar{\lambda}_{2}-
\lambda_{2}\bar{\lambda}_{0}-\lambda_{3}\bar{\lambda}_{1}-
\lambda_{1}\bar{\lambda}_{3})+g_{0}(\lambda_{2}\bar{\lambda}_{3}-
\lambda_{3}\bar{\lambda}_{2}-\lambda_{0}\bar{\lambda}_{1}-\lambda_{1}\bar{\lambda}_{0})+\\\nonumber
&&\bar{g}_{0}(-\bar{\lambda}_{0}\bar{\lambda}_{1}+
\lambda_{1}\lambda_{0}-\bar{\lambda}_{2}\bar{\lambda}_{3}+
\lambda_{2}\lambda_{3})+\bar{g}_{2}(\bar{\lambda}_{0}\bar{\lambda}_{3}+
\lambda_{3}\lambda_{0}+\bar{\lambda}_{2}\bar{\lambda}_{1}+\lambda_{2}\lambda_{1})+\\\nonumber
&&\bar{g}_{3}(-\bar{\lambda}_{0}\bar{\lambda}_{2}-
\lambda_{2}\lambda_{0}+\bar{\lambda}_{3}\bar{\lambda}_{1}+\lambda_{3}\lambda_{1})]   +\\\nonumber
&&\Phi_{\bar{1}\bar{1}}(\bar{\lambda}_{1}\lambda_{1}\lambda_{2}
\bar{\lambda}_{2}+\bar{\lambda}_{1}\lambda_{1}
\lambda_{3}\bar{\lambda}_{3}+\bar{\lambda}_{0}\lambda_{1}
\bar{\lambda}_{2}\lambda_{3}+\bar{\lambda}_{0}\lambda_{2}
\bar{\lambda}_{3}\lambda_{1}+\bar{\lambda}_{0}\lambda_{1}\bar{\lambda}_{2}\lambda_{3}+\\\nonumber
&&\bar{\lambda}_{0}\lambda_{3}\bar{\lambda}_{1}\lambda_{2}-
\bar{\lambda}_{1}\bar{\lambda}_{2}\bar{\lambda}_{3}\bar{\lambda}_{0})         +  \\\nonumber
&&\Phi_{11}(\bar{\lambda}_{0}\lambda_{0}\lambda_{1}
\bar{\lambda}_{1}+\bar{\lambda}_{0}
\lambda_{2}\lambda_{3}\bar{\lambda}_{1}+\bar{\lambda}_{1}\lambda_{1}\lambda_{2}\bar{\lambda}_{2}+
\bar{\lambda}_{1}\lambda_{1}\lambda_{3}
\bar{\lambda}_{3}+\lambda_{0}\lambda_{2}\bar{\lambda}_{3}
\bar{\lambda}_{1}+\lambda_{0}\lambda_{3}\bar{\lambda}_{1}\bar{\lambda}_{2}\\\nonumber
&&-\lambda_{1}\lambda_{2}\lambda_{3}\lambda_{0})+\\\nonumber
&&\Phi_{22}(\bar{\lambda}_{0}\lambda_{0}\lambda_{2}\bar{\lambda}_{2}+
\bar{\lambda}_{0}\lambda_{3}\lambda_{1}
\bar{\lambda}_{2}+\bar{\lambda}_{2}\lambda_{2}\lambda_{1}\bar{\lambda}_{1}+
\bar{\lambda}_{2}\lambda_{2}\lambda_{3}\bar{\lambda}_{3}+
\lambda_{0}\lambda_{1}\bar{\lambda}_{2}\bar{\lambda}_{3}+\lambda_{0}\lambda_{3}\bar{\lambda}_{1}\bar{\lambda}_{2}-\\\nonumber
&&-\lambda_{1}\lambda_{2}\lambda_{3}\lambda_{0})+\\\nonumber
        &&\Phi_{33}(\bar{\lambda}_{0}\lambda_{0}\lambda_{3}\bar{\lambda}_{3}+
\bar{\lambda}_{0}\lambda_{1}\lambda_{2}\bar{\lambda}_{3}+
\bar{\lambda}_{3}\lambda_{3}\lambda_{1}\bar{\lambda}_{1}+\bar{\lambda}_{3}\lambda_{3}\lambda_{2}\bar{\lambda}_{2}+
\lambda_{0}\lambda_{1}\bar{\lambda}_{2}\bar{\lambda}_{3}+\lambda_{0}\lambda_{2}\bar{\lambda}_{3}\bar{\lambda}_{1}\\\nonumber
&&-\lambda_{1}\lambda_{2}\lambda_{3}\lambda_{0})+\\\nonumber
 &&\Phi_{\bar{1}1}(\lambda_{1}\lambda_{0}\lambda_{2}\bar{\lambda}_{3}-
\bar{\lambda}_{0}\bar{\lambda}_{1}\bar{\lambda}_{2}\lambda_{3}+
\lambda_{0}\lambda_{3}\lambda_{1}\bar{\lambda}_{2}+\bar{\lambda}_{0}\bar{\lambda}_{3}\bar{\lambda}_{1}\lambda_{2}+
\lambda_{0}\lambda_{2}\lambda_{3}\bar{\lambda}_{1}-\bar{\lambda}_{0}\bar{\lambda}_{2}\bar{\lambda}_{3}\lambda_{1}\\\nonumber
&&-\lambda_{1}\lambda_{2}\lambda_{3}
\bar{\lambda}_{0}-\bar{\lambda}_{1}\bar{\lambda}_{2}\bar{\lambda}_{3}\lambda_{0})+\\\nonumber
&&\Phi_{\bar{1}2}(\lambda_{0}\lambda_{2}\lambda_{3}\bar{\lambda}_{2}+
\bar{\lambda}_{2}\bar{\lambda}_{1}\lambda_{0}\bar{\lambda}_{0}+\bar{\lambda}_{0}\bar{\lambda}_{2}\bar{\lambda}_{3}\lambda_{2}+
\lambda_{0}\lambda_{3}\lambda_{1}\bar{\lambda}_{1}-\bar{\lambda}_{0}\bar{\lambda}_{3}\bar{\lambda}_{1}\lambda_{1}-
\lambda_{1}\lambda_{2}\lambda_{3}\bar{\lambda}_{3}\\\nonumber
&&-\bar{\lambda}_{1}\bar{\lambda}_{2}\lambda_{3}\bar{\lambda}_{3})                     +\\\nonumber
&&\Phi_{\bar{1}3}(\lambda_{0}\lambda_{2}\lambda_{3}
\bar{\lambda}_{3}+\bar{\lambda}_{3}\bar{\lambda}_{1}\lambda_{0}
\bar{\lambda}_{0}+\bar{\lambda}_{0}\bar{\lambda}_{2}
\bar{\lambda}_{3}\lambda_{3}+\lambda_{0}\lambda_{1}
\lambda_{2}\bar{\lambda}_{1}-\bar{\lambda}_{0}
\bar{\lambda}_{1}\bar{\lambda}_{2}\lambda_{1}-\lambda_{1}\lambda_{3}\lambda_{2}\bar{\lambda}_{2}\\\nonumber
&&-\bar{\lambda}_{1}\bar{\lambda}_{3}\lambda_{2}\bar{\lambda}_{2})                     +\\\nonumber
&&\Phi_{12}(\bar{\lambda}_{0}\lambda_{0}\lambda_{2}
\bar{\lambda}_{1}-\lambda_{0}
\bar{\lambda}_{0}\lambda_{1}\bar{\lambda}_{2}+\bar{\lambda}_{0}\lambda_{2}\lambda_{3}\bar{\lambda}_{2}-\lambda_{0}
\bar{\lambda}_{2}\bar{\lambda}_{3}\lambda_{2}+
\bar{\lambda}_{0}\lambda_{3}\lambda_{1}\bar{\lambda}_{1}-\lambda_{0}\bar{\lambda}_{3}\bar{\lambda}_{1}\lambda_{1}\\\nonumber
&&+\bar{\lambda}_{1}
\lambda_{2}\lambda_{3}\bar{\lambda}_{3}+\bar{\lambda}_{2}\lambda_{1}\lambda_{3}\bar{\lambda}_{3})                   +  \\\nonumber
&&\Phi_{13}(\bar{\lambda}_{0}\lambda_{0}\lambda_{3}\bar{\lambda}_{1}-
\lambda_{0}\bar{\lambda}_{0}\lambda_{1}\bar{\lambda}_{3}+\bar{\lambda}_{0}\lambda_{1}\lambda_{2}\bar{\lambda}_{1}-\lambda_{0}
\bar{\lambda}_{1}\bar{\lambda}_{2}\lambda_{1}+\bar{\lambda}_{0}
\lambda_{2}\lambda_{3}\bar{\lambda}_{3}-\lambda_{0}\bar{\lambda}_{2}\bar{\lambda}_{3}\lambda_{3}\\\nonumber
&&+\bar{\lambda}_{1}\lambda_{3}\lambda_{2}
\bar{\lambda}_{2}+\bar{\lambda}_{3}\lambda_{1}\lambda_{2}\bar{\lambda}_{2})                    + \\\nonumber
&&\Phi_{23}(\bar{\lambda}_{0}\lambda_{0}
\lambda_{3}\bar{\lambda}_{2}-\lambda_{0}\bar{\lambda}_{0}
\lambda_{2}\bar{\lambda}_{3}+\bar{\lambda}_{0}\lambda_{1}\lambda_{2}\bar{\lambda}_{2}-\lambda_{0}
\bar{\lambda}_{1}\bar{\lambda}_{2}\lambda_{2}+
\bar{\lambda}_{0}\lambda_{3}\lambda_{1}\bar{\lambda}_{3}-\lambda_{0}\bar{\lambda}_{3}\bar{\lambda}_{1}\lambda_{3}\\\nonumber
&&+\bar{\lambda}_{2}
\lambda_{3}\lambda_{1}\bar{\lambda}_{1}+\bar{\lambda}_{3}\lambda_{2}\lambda_{1}\bar{\lambda}_{1})                   +  \\\nonumber
&&\Omega(\dot{\upsilon}_{1}
\dot{\bar{\upsilon}}_{1}+g_{0}
\bar{g}_{0}+\dot{\upsilon}_{2}\bar{g}_{2}+
\dot{\upsilon}_{3}\bar{g}_{3}+\lambda_{0}
\dot{\bar{\lambda}}_{0}+\lambda_{1}\dot{\bar{\lambda}}_{1}+
\lambda_{2}\dot{\bar{\lambda}}_{2}+\lambda_{3}\dot{\bar{\lambda}}_{3}) +\\\nonumber
&&\Omega_{1}(\dot{\upsilon}_{1}\lambda_{1}\bar{\lambda}_{1}+
\dot{\upsilon}_{2}\lambda_{2}\bar{\lambda}_{1}+\dot{\upsilon}_{3}\lambda_{3}\bar{\lambda}_{1}+
g_{0}\lambda_{0}\bar{\lambda}_{1}-\bar{g}_{0}\lambda_{2}\lambda_{3}+
\bar{g}_{2}\lambda_{0}\lambda_{3}-\bar{g}_{3}\lambda_{0}\lambda_{2}) +\\\nonumber
&&\Omega_{2}(\dot{\upsilon}_{1}\lambda_{1}\bar{\lambda}_{2}+
\dot{\upsilon}_{2}\lambda_{2}\bar{\lambda}_{2}+\dot{\upsilon}_{3}\lambda_{3}\bar{\lambda}_{2}-\dot{\bar{\upsilon}}_{1}
\lambda_{0}\lambda_{3}-
g_{0}\bar{\lambda}_{2}\lambda_{0}-\bar{g}_{0}\lambda_{3}\lambda_{1}+\bar{g}_{3}\lambda_{0}\lambda_{1})+ \\\nonumber
&&\Omega_{3}(\dot{\upsilon}_{1}\lambda_{1}
\bar{\lambda}_{3}+\dot{\upsilon}_{2}
\lambda_{2}\bar{\lambda}_{3}+\dot{\upsilon}_{3}\lambda_{3}\bar{\lambda}_{3}+\dot{\bar{\upsilon}}_{1}\lambda_{0}\lambda_{2}-g_{0}
\bar{\lambda}_{3}\lambda_{0}-\bar{g}_{0}\lambda_{1}\lambda_{2}-\bar{g}_{2}\lambda_{0}\lambda_{1}) +\\\nonumber
&&\Omega_{\bar{1}}(\dot{\upsilon}_{1}(\lambda_{2}
\bar{\lambda}_{2}+\lambda_{3}\bar{\lambda}_{3})+
\dot{\upsilon}_{2}\bar{\lambda}_{0}
\bar{\lambda}_{3}-\dot{\upsilon}_{3}\bar{\lambda}_{0}\bar{\lambda}_{2}-g_{0}
\bar{\lambda}_{2}\bar{\lambda}_{3}+\bar{g}_{0}\bar{\lambda}_{0}\lambda_{1}+
\bar{g}_{2}\bar{\lambda}_{2}\lambda_{1}+\bar{g}_{3}\bar{\lambda}_{3}\lambda_{1})+ \\\nonumber
&&\Omega_{\bar{1}1}(\lambda_{0}
\bar{\lambda}_{0}\lambda_{1}\bar{\lambda}_{1}+
\lambda_{1}\bar{\lambda}_{1}\lambda_{2}\bar{\lambda}_{2}+\lambda_{1}\bar{\lambda}_{1}\lambda_{3}\bar{\lambda}_{3})+\\\nonumber
&&\Omega_{\bar{1}2}(\lambda_{0}\bar{\lambda}_{0}\lambda_{1}\bar{\lambda}_{2}+\lambda_{1}\bar{\lambda}_{2}\lambda_{3}\bar{\lambda}_{3})
+\\\nonumber
&&\Omega_{\bar{1}3}(\lambda_{0}\bar{\lambda}_{0}\lambda_{1}\bar{\lambda}_{3}+\lambda_{1}\bar{\lambda}_{3}\lambda_{2}\bar{\lambda}_{2})
+\\\nonumber
&&\Omega_{12}(\lambda_{2}\lambda_{0}\lambda_{3}\bar{\lambda}_{2}-\lambda_{0}\lambda_{3}\lambda_{1}\bar{\lambda}_{1})+\\\nonumber
&&\Omega_{13}(\lambda_{1}\lambda_{0}\lambda_{2}\bar{\lambda}_{1}-\lambda_{0}\lambda_{2}\lambda_{3}\bar{\lambda}_{3})+\\\nonumber
&&\Omega_{23}(\lambda_{1}\lambda_{0}\lambda_{2}\bar{\lambda}_{2}-\lambda_{0}\lambda_{3}\lambda_{1}\bar{\lambda}_{3}),\\\nonumber
\end{eqnarray}
where $F(\upsilon_{1},\upsilon_{2},\upsilon_{3},\bar{\upsilon}_{1})$ is the unconstrained prepotential, while
\bea
\Omega&=&\Box F= \partial_{11}F+\partial_{22}F+\partial_{33}F+\partial_{\bar{1}\bar{1}}F,\nonumber\\
\Phi&=&\partial_{1\bar{1}}F
\eea
and the partial derivative of $\Omega$ ($\Phi$)  w.r.t. $\upsilon_i$, $\bar{\upsilon}_1$
is expressed as $\Omega_i$, $\Omega_{\bar 1}$ ($\Phi_i$, ${\Phi}_{\bar 1}$), respectively.
\par
Similarly to the results of \cite{grt}, the constraints $\Box\Phi=0$ and $ \Omega=0$ arise as a consequence of
imposing an extra
invariance under an ${\cal N}=5$-Extended Supersymmetry (under such constraints the resulting off-shell action
is also automatically
${\cal N}=8$-invariant).\par
An inequivalent ${\cal N}=4$-invariant action is obtained by applying the Construction I to the supermultiplet $(4,8,4)_D$.\par
For $(2,8,6)_B$ the associated Lagrangian is
\begin{eqnarray}\label{286B}
{\cal L}&=&\Phi(\dot{\upsilon}_{0}^{2}+
\dot{\bar{\upsilon}}_{0}^{2}+g_{i}^{2}+
\bar{g}_{i}^{2}+\dot{\lambda}_{0}\lambda_{0}+\dot{\bar{\lambda}}_{0}\bar{\lambda}_{0}+\dot{\lambda}_{i}\lambda_{i}+
\dot{\bar{\lambda}}_{i}\bar{\lambda}_{i})+\\  \nonumber
&&\Phi_{0}[\dot{\bar{\upsilon}}_{0}(\bar{\lambda}_{i}
\lambda_{i}+\lambda_{0}\bar{\lambda}_{0})+g_{i}
(\bar{\lambda}_{0}\bar{\lambda}_{i}-\lambda_{i}\lambda_{0})+\bar{g}_{i}(\lambda_{0}\bar{\lambda}_{i}+\lambda_{i}
\bar{\lambda}_{0})\\\nonumber
&&-\frac{\varepsilon_{ijk}}{2}g_{i}(\bar{\lambda}_{j}
\bar{\lambda}_{k}-\lambda_{j}\lambda_{k})-\varepsilon_{ijk}\lambda_{i}\bar{\lambda}_{j}\bar{g}_{k}]+\\\nonumber
&&\Phi_{\bar{0}}[\dot{\upsilon}_{0}(\lambda_{i}\bar{\lambda}_{i}+
\bar{\lambda}_{0}\lambda_{0})+\bar{g}_{i}(\bar{\lambda}_{0}\bar{\lambda}_{i}-\lambda_{i}\lambda_{0})-g_{i}
(\lambda_{0}\bar{\lambda}_{i}+\lambda_{i}\bar{\lambda}_{0})\\\nonumber
&&+\frac{\varepsilon_{ijk}}{2}\bar{g}_{i}
(\bar{\lambda}_{j}\bar{\lambda}_{k}-\lambda_{j}\lambda_{k})-\varepsilon_{ijk}\lambda_{i}\bar{\lambda}_{j}g_{k}]+\\\nonumber
&&\Phi_{00}\frac{\varepsilon_{ijk}}{6}
[3\bar{\lambda}_{0}\lambda_{i}\bar{\lambda}_{j}\lambda_{k}+\lambda_{0}\lambda_{i}\lambda_{j}\lambda_{k}]+\\\nonumber
&&\Phi_{\bar{0}\bar{0}}
\frac{\varepsilon_{ijk}}{6}[3\lambda_{0}
\lambda_{i}\bar{\lambda}_{j}\bar{\lambda}_{k}+\bar{\lambda}_{0}\bar{\lambda}_{i}\bar{\lambda}_{j}\bar{\lambda}_{k}]\\\nonumber
&&-(\Phi_{\bar{0}\bar{0}}+\Phi_{00})\lambda_{0}\bar{\lambda}_{0}\lambda_{i}\bar{\lambda}_{i}+\\\nonumber
&&\Phi_{0\bar{0}}\frac{\varepsilon_{ijk}}{6}
[\lambda_{i}\lambda_{j}\lambda_{k}\bar{\lambda}_{0}+
\bar{\lambda}_{i}\bar{\lambda}_{j}\bar{\lambda}_{k}\lambda_{0}-3(\lambda_{0}\lambda_{i}\lambda_{j}\bar{\lambda}_{k}+
\bar{\lambda}_{0}\bar{\lambda}_{i}\bar{\lambda}_{j}\lambda_{k})]+\\\nonumber
&&\Omega(\dot{\upsilon}_{0}\dot{\bar{\upsilon}}_{0}+
\lambda_{0}\dot{\bar{\lambda}}_{0}+\lambda_{i}\dot{\bar{\lambda}}_{i}+g_{i}\bar{g}_{i})\\\nonumber
&&-\Omega_{0}(\dot{\upsilon}_{0}\bar{\lambda}_{0}\lambda_{0}+g_{i}\bar{\lambda}_{0}\lambda_{i})+\\\nonumber
&&\Omega_{\bar{0}}(\dot{\bar{\upsilon}}_{0}\bar{\lambda}_{i}\lambda_{i}+\bar{g}_{i}\bar{\lambda}_{i}\lambda_{0})+\\\nonumber
&&\Omega_{0\bar{0}}\lambda_{0}\bar{\lambda}_{0}\lambda_{i}\bar{\lambda}_{i}\\\nonumber
&&-\frac{\varepsilon_{ijk}}{6}
(\Omega_{00}\bar{\lambda}_{0}
\lambda_{i}\lambda_{j}\lambda_{k}+\Omega_{\bar{0}\bar{0}}\bar{\lambda}_{i}\bar{\lambda}_{j}\bar{\lambda}_{k}\lambda_{0}),
\end{eqnarray}
where now we have
\bea
\Omega&=&\partial_{00}F+\partial_{\bar{0}\bar{0}}F,\nonumber\\
\Phi&=&\partial_{0\bar{0}}F.
\eea
For $(2,8,6)_b$ the associated Lagrangian is
\begin{eqnarray}\label{286b}
{\cal L}&=&\Gamma(\dot{\upsilon}_{0}^{2}+g_{i}^{2}+\dot{\lambda}_{0}
\lambda_{0}+\dot{\lambda}_{i}\lambda_{i})-
\bar{\Gamma}(\dot{\bar{\upsilon}}_{0}^{2}+
\bar{g}_{i}^{2}+\dot{\bar{\lambda}}_{0}\bar{\lambda}_{0}+\dot{\bar{\lambda}}_{i}\bar{\lambda}_{i})+\\\nonumber
&&\bar{\Gamma}_{0}\dot{\bar{\upsilon}}_{0}
(\lambda_{i}\bar{\lambda}_{i}+\lambda_{0}
\bar{\lambda}_{0})-g_{i}(\bar{\Gamma}_{0}\bar{\lambda}_{0}\bar{\lambda}_{i}+\Gamma_{0}\lambda_{i}\lambda_{0})-
\bar{\Gamma}_{0}\bar{g}_{i}(\lambda_{0}\bar{\lambda}_{i}+\lambda_{i}\bar{\lambda}_{0})  + \\\nonumber
&&g_{i}\frac{\varepsilon_{ijk}}{2}
(\bar{\Gamma}_{0}\bar{\lambda}_{i}\bar{\lambda}_{j}+
\Gamma_{0}\lambda_{j}\lambda_{k})+\bar{\Gamma}_{0}\varepsilon_{ijk}\lambda_{i}\bar{\lambda}_{j}\bar{g}_{k}+
\\\nonumber
&&\Gamma_{\bar{0}}\dot{\upsilon}_{0}(\lambda_{i}
\bar{\lambda}_{i}+\lambda_{0}\bar{\lambda}_{0})-
\bar{g}_{i}(\bar{\Gamma}_{\bar{0}}\bar{\lambda}_{0}
\bar{\lambda}_{i}+\Gamma_{\bar{0}}\lambda_{i}\lambda_{0})-\Gamma_{\bar{0}}g_{i}(\lambda_{0}\bar{\lambda}_{i}+\lambda_{i}
\bar{\lambda}_{0})
\\\nonumber
&&-\bar{g}_{i}\frac{\varepsilon_{ijk}}{2}(\bar{\Gamma}_{\bar{0}}
\bar{\lambda}_{i}\bar{\lambda}_{j}+\Gamma_{\bar{0}}\lambda_{j}\lambda_{k})-\Gamma_{\bar{0}}\varepsilon_{ijk}\lambda_{i}
\bar{\lambda}_{j}g_{k}
\\\nonumber
&&-\frac{\varepsilon_{ijk}}{6}
(3\bar{\Gamma}_{00}\bar{\lambda}_{0}\lambda_{i}\bar{\lambda}_{j}\lambda_{k}-\Gamma_{00}\lambda_{0}\lambda_{i}\lambda_{j}
\lambda_{k})+
\\\nonumber
&&\frac{\varepsilon_{ijk}}{6}(3\Gamma_{\bar{0}\bar{0}}
\lambda_{0}\bar{\lambda}_{i}\lambda_{j}\bar{\lambda}_{k}-
\bar{\Gamma}_{\bar{0}\bar{0}}\bar{\lambda}_{0}\bar{\lambda}_{i}\bar{\lambda}_{j}\bar{\lambda}_{k})
\\\nonumber
&&-\Gamma_{0\bar{0}}\frac{\varepsilon_{ijk}}{6}
(\lambda_{i}\lambda_{j}\lambda_{k}\bar{\lambda}_{0}+3\lambda_{0}\lambda_{i}
\lambda_{j}\bar{\lambda}_{k})+\bar{\Gamma}_{0\bar{0}}\frac{\varepsilon_{ijk}}{6}
(\bar{\lambda}_{i}\bar{\lambda}_{j}\bar{\lambda}_{k}\lambda_{0}+3\bar{\lambda}_{0}\bar{\lambda}_{i}\bar{\lambda}_{j}\lambda_{k}),
\end{eqnarray}
with
\bea
\Gamma&=&\partial_{00}F(\upsilon_{0},\bar{\upsilon}_{0}),\nonumber\\ \bar{\Gamma}&=&\partial_{\bar{0}\bar{0}}F(\upsilon_{0},
\bar{\upsilon}_{0}).
\eea
Therefore it turns out, comparing (\ref{286B}) with (\ref{286b}), that the two inequivalent ${\cal N}=4$ supermultiplets
$(2,8,6)_B$ and $(2,8,6)_b$ produce inequivalent (for generic values of the functions $\Omega, \Phi, \Gamma, \bar{\Gamma}$) ${\cal N}=4$
off-shell invariant actions.\par
The extra-requirement of an ${\cal N}=5$-invariance produces the constraints $\Box \Phi=0$, $\Omega=0$ for the $(2,8,6)_B$ case and
$\Gamma=-\bar{\Gamma}$ (implying $\Box\Gamma=0$) for the $(2,8,6)_b$ case. The resulting constrained Lagrangians coincide and the action possesses an
overall ${\cal N}=8$ invariance.\par
It is interesting to present an explicit example of ${\cal N}=4$-invariant, first-order action derived from Construction I.
For the $(4,8,4)_{B}$ non-minimal supermultiplet with connectivity symbol $4_{3}+4_{1}$ and
component fields $(\upsilon_{0},\upsilon_{i};\lambda_{0},\lambda_{i},\bar{\lambda}_{0},\bar{\lambda}_{i};\bar{g}_{0},\bar{g}_{i})$
($i=1,2,3$),
we have that the associated Lagrangian is
\begin{eqnarray}\label{484BI}
{\cal L}&=&\Omega(\dot{\upsilon}_{0}\bar{g}_{0}+\dot{\upsilon}_{i}\bar{g}_{i})+
\Omega(\lambda_{0}\dot{\bar{\lambda}}_{0}+\lambda_{k}\dot{\bar{\lambda}}_{k})+\\\nonumber
&&\varepsilon_{ijk}\Omega_{j}\bar{g}_{k}\lambda_{0}
\lambda_{i}-\Omega_{i}\dot{\upsilon}_{0}
\bar{\lambda}_{i}\lambda_{0}-\Omega_{0}\dot{\upsilon}_{i}\bar{\lambda}_{0}\lambda_{i}+\\  \nonumber
&&\frac{\varepsilon_{ijk}}{2}(\Omega_{0}\bar{g}_{k}-
\Omega_{k}\bar{g}_{0})\lambda_{i}\lambda_{j}-
\Omega_{j}\dot{\upsilon}_{i}\bar{\lambda}_{j}\lambda_{i}-\Omega_{0}\dot{\upsilon}_{0}\bar{\lambda}_{0}\lambda_{0}\\\nonumber
&&-\frac{1}{2}\varepsilon_{ijk}
\Omega_{0k}\lambda_{i}\lambda_{j}\lambda_{0}\bar{\lambda}_{0}-
\frac{\varepsilon_{ijk}}{2}\delta_{pq}\Omega_{pk}\lambda_{0}\lambda_{i}\lambda_{j}\bar{\lambda}_{q}\\\nonumber
&&-\frac{\varepsilon_{ijk}}{6}(\Omega_{00}\bar{\lambda}_{0}-\Omega_{0p}\lambda_{p})\lambda_{i}\lambda_{j}\lambda_{k},\\\nonumber
\end{eqnarray}
where
\begin{equation}
\Omega=\partial_{00}F+\partial_{ii}F.
\end{equation}
In the next Section we show how, for this one and the other supermultiplets entering (\ref{firstordermult}),
by applying Construction II, we obtain a second-order, ${\cal N}=4$-invariant action expressed in terms of
a constrained
prepotential.

\section{${\cal N}=4$-invariant $\sigma$-models with a constrained prepotential}

The application of Construction II to the non-minimal supermultiplet $(4,8,4)_B$ induces an ${\cal N}=4$-invariant
theory, obtained from a second-order Lagrangian, which is expressed in terms of the two independent functions, derived from the prepotential $W$ entering
(\ref{Wprep}),
$\Phi(\upsilon_{0},\upsilon_{i})$ and $\Omega(\upsilon_{0},\upsilon_{i})$. The ${\cal N}=4$-invariance is however
recovered if and only if the constraint
\bea
\Phi_{00}+\Phi_{ii}&=&0
\eea
is satisfied. This feature is not present for the ${\cal N}=4$ actions derived from the (\ref{secondordermult}) supermultiplets. \par
The Lagrangian is explicitly given by
\begin{eqnarray}\label{484B}
{\cal L}&=&\Phi(\dot{\upsilon}_{0}^{2}+
\dot{\upsilon}_{i}^{2}+\bar{g}_{0}^{2}+\bar{g}_{i}^{2}+
\dot{\lambda}_{0}\lambda_{0}+\dot{\bar{\lambda}}_{0}\bar{\lambda}_{0}+\dot{\lambda}_{i}\lambda_{i}+\dot{\bar{\lambda}}_{i}
\bar{\lambda}_{i})+\\ \nonumber
&&\varepsilon_{ijk}\Phi_{k}\dot{\upsilon}_{j}
(\bar{\lambda}_{0}\bar{\lambda}_{i}+\lambda_{i}\lambda_{0})+(\Phi_{0}\dot{\upsilon}_{i}-\Phi_{i}\dot{\upsilon}_{0})(\bar{\lambda}_{0}\bar{\lambda}_{i}-\lambda_{i}\lambda_{0})\\\nonumber
&&-\varepsilon_{ijk}\Phi_{j}\bar{g}_{k}
(\lambda_{0}\bar{\lambda}_{i}-\lambda_{i}\bar{\lambda}_{0})+
(\Phi_{0}\bar{g}_{i}+\Phi_{i}\bar{g}_{0})(\lambda_{0}\bar{\lambda}_{i}+\lambda_{i}\bar{\lambda}_{0})+\\\nonumber
&&\frac{\varepsilon_{ijk}}{2}
(\Phi_{i}\dot{\upsilon}_{0}-\Phi_{0}
\dot{\upsilon}_{i})(\bar{\lambda}_{j}\bar{\lambda}_{k}-\lambda_{j}\lambda_{k})+\frac{1}{2}(\Phi_{j}\dot{\upsilon}_{k}-\Phi_{k}
\dot{\upsilon}_{j})(\bar{\lambda}_{j}\bar{\lambda}_{k}+\lambda_{j}\lambda_{k})+\\\nonumber
&&(\Phi_{0}\bar{g}_{0}+\Phi_{j}\bar{g}_{j})
(\bar{\lambda}_{0}\lambda_{0}+
\bar{\lambda}_{k}\lambda_{k})-\Phi_{i}\bar{g}_{j}
(\bar{\lambda}_{i}\lambda_{j}-\lambda_{i}\bar{\lambda}_{j})-\Phi_{0}\bar{g}_{0}(\bar{\lambda}_{0}\lambda_{0}-\lambda_{0}
\bar{\lambda}_{0})+\\\nonumber
&&\varepsilon_{ijk}\lambda_{i}\bar{\lambda}_{j}
(\Phi_{k}\bar{g}_{0}-\Phi_{0}\bar{g}_{k})+(\varepsilon_{ijk}\Phi_{0k}-\Phi_{ij})\lambda_{0}\bar{\lambda}_{0}\lambda_{i}
\bar{\lambda}_{j}+\\\nonumber
&&\Phi_{0j}(\bar{\lambda}_{0}\lambda_{j}-
\lambda_{0}\bar{\lambda}_{j})\lambda_{i}\bar{\lambda}_{i}+
\frac{1}{2}\varepsilon_{ijk}\delta_{pq}\Phi_{kp}(\bar{\lambda}_{0}\lambda_{i}\lambda_{j}\bar{\lambda}_{q}-\lambda_{0}
\bar{\lambda}_{i}\bar{\lambda}_{j}\lambda_{q})+\\\nonumber
&&\Phi_{jk}\bar{\lambda}_{j}\lambda_{k}\lambda_{i}
\bar{\lambda}_{i}-\frac{1}{2}\varepsilon_{ijk}\delta_{pq}\Phi_{0p}\lambda_{i}\bar{\lambda}_{j}\lambda_{k}\bar{\lambda}_{q}\\\nonumber
&&-\Phi_{00}\lambda_{0}\bar{\lambda}_{0}\lambda_{i}
\bar{\lambda}_{i}+\frac{\varepsilon_{ijk}}{2}
[\Phi_{pp}(\lambda_{0}\lambda_{i}\bar{\lambda}_{j}\bar{\lambda}_{k})+\Phi_{00}(\bar{\lambda}_{0}
\lambda_{i}\bar{\lambda}_{j}\lambda_{k})]\\\nonumber
&&-\frac{\varepsilon_{ijk}}{6}[(\Phi_{00}+\Phi_{pp})\lambda_{i}\lambda_{j}\lambda_{k}\lambda_{0}]+\\\nonumber
&&\Omega(\dot{\upsilon}_{0}\bar{g}_{0}+\dot{\upsilon}_{i}\bar{g}_{i})+
\Omega(\lambda_{0}\dot{\bar{\lambda}}_{0}+\lambda_{k}\dot{\bar{\lambda}}_{k})+\\\nonumber
&&\varepsilon_{ijk}\Omega_{j}\bar{g}_{k}\lambda_{0}\lambda_{i}-\Omega_{i}
\dot{\upsilon}_{0}\bar{\lambda}_{i}\lambda_{0}-\Omega_{0}\dot{\upsilon}_{i}\bar{\lambda}_{0}\lambda_{i}+\\  \nonumber
&&\frac{\varepsilon_{ijk}}{2}(\Omega_{0}\bar{g}_{k}-
\Omega_{k}\bar{g}_{0})\lambda_{i}\lambda_{j}-\Omega_{j}
\dot{\upsilon}_{i}\bar{\lambda}_{j}\lambda_{i}-\Omega_{0}\dot{\upsilon}_{0}\bar{\lambda}_{0}\lambda_{0}\\\nonumber
&&-\frac{1}{2}\varepsilon_{ijk}\Omega_{0k}\lambda_{i}\lambda_{j}
\lambda_{0}\bar{\lambda}_{0}-\frac{\varepsilon_{ijk}}{2}\delta_{pq}\Omega_{pk}\lambda_{0}\lambda_{i}\lambda_{j}
\bar{\lambda}_{q}\\\nonumber
&&-\frac{\varepsilon_{ijk}}{6}(\Omega_{00}\bar{\lambda}_{0}-\Omega_{0p}\lambda_{p})\lambda_{i}\lambda_{j}\lambda_{k}\\\nonumber
\end{eqnarray}
Requiring an extra, ${\cal N}=5$-invariance for the action implies the further constraint $\Omega=0$. The resulting action is automatically ${\cal N}=8$-invariant and coincides with the ${\cal N}=5$ constrained action derived from $(4,8,4)_C$.\par
All actions obtained via Construction II from the supermultiplets entering (\ref{firstordermult}) share the same features.
In the case of the
supermultiplet $(1,8,7)_{red}$, described by the fields $(\upsilon_{0};\lambda_{0},
\lambda_{i},\bar{\lambda}_{0},\bar{\lambda}_{i};g_{i},\bar{g}_{0},
\bar{g}_{i})$, the associated Lagrangian is
\begin{eqnarray}\label{187}
{\cal L}&=&\Phi(\dot{\upsilon}_{0}^{2}+\bar{g}_{0}^{2}+g_{i}^{2}+\bar{g}_{i}^{2}+\dot{\lambda}_{0}\lambda_{0}+
\dot{\bar{\lambda}}_{0}\bar{\lambda}_{0}+\dot{\lambda}_{i}\lambda_{i}+\dot{\bar{\lambda}}_{i}
\bar{\lambda}_{i})\\\nonumber
&&\Phi_{0}[\bar{g}_{0}(\lambda_{0}\bar{\lambda}_{0}-
\lambda_{i}\bar{\lambda}_{i})+g_{i}(\bar{\lambda}_{0}\bar{\lambda}_{i}-
\lambda_{i}\lambda_{0})+\bar{g}_{i}(\lambda_{0}\bar{\lambda}_{i}+\lambda_{i}\bar{\lambda}_{0})-\\\nonumber
&&-\varepsilon_{ijk}(\bar{g}_{i}
\bar{\lambda}_{j}\lambda_{k}+\frac{g_{i}}{2}(\bar{\lambda}_{j}
\bar{\lambda}_{k}-\lambda_{j}\lambda_{k}))]+\\\nonumber
&&\Omega(\dot{\upsilon}_{0}\bar{g}_{0}+
\lambda_{0}\dot{\bar{\lambda}}_{0}+\lambda_{i}\dot{\bar{\lambda}}_{i}+g_{i}\bar{g}_{i})+\\\nonumber
&&\Omega_{0}[(\dot{\upsilon}_{0}\lambda_{0}+
g_{i}\lambda_{i})\bar{\lambda}_{0}+\frac{1}{2}\varepsilon_{ijk}\bar{g}_{i}\lambda_{j}\lambda_{k}]+\\\nonumber
&&\Omega_{00}\frac{1}{6}\varepsilon_{ijk}\lambda_{i}\lambda_{j}\lambda_{k}\bar{\lambda}_{0},
\end{eqnarray}
where $\Phi$, $\Omega$ are fuctions of $\upsilon_0$. The ${\cal N}=4$-invariance requires
\begin{equation}
\Phi_{00}=0.
\end{equation}
Implementing the ${\cal N}=5$-invariance gives the further constraint $\Omega=0$ (again, the resulting action is automatically fully
${\cal N}=8$-invariant).\par
The ${\cal N}=8$ model based on the supermultiplet of field content $(1,8,7)$ was first
obtained in \cite{krt}. Unlike the present, more general
construction, the  action was derived through an ``${\cal N}=8$ covariantization Ansatz"
which cannot be applied neither to deduce the ${\cal N}=4$-invariant action, nor to obtain the invariant actions
for the other supermultiplets entering (\ref{firstordermult}).

\section{Conclusions}

We have already sketched in the Introduction the structure of our paper and presented its main results.
Here we limit ourselves to make some comments and outline the implications of our work. \par
It is worth stressing the fact that we obtained here for the first time evidence that supermultiplets,
sharing the same field content but differing in connectivity symbol, can induce inequivalent supersymmetric-invariant
actions (one should compare, e.g., the actions given in formulas (\ref{484C}) and (\ref{484B})). It was known,
from the analysis of \cite{{dfghil},{dfghil2},{kt},{kt2}}, that inequivalent representations, discriminated by
their respective connectivity symbol, can be found. On the other hand, so far, no dynamical characterization
was associated to the connectivity symbol. In \cite{grt} the ${\cal N}=5$-supersymmetric off-shell invariant
actions, induced with respect to inequivalent ${\cal N}=5$ supermultiplets of a given field content, were
proven to coincide and possess an overall ${\cal N}=8$ supersymmetry invariance. The crucial feature here is
the fact that inequivalent ${\cal N}=4$ off-shell invariant actions are induced by inequivalent
{\em non-minimal} ${\cal N}=4$ linear supermultiplets (with the same field content).\par
We have to spend, as promised, some words on the concept of {\em oxidation}. It is a pun, employed in
superstring/M-theory literature, see \cite{top}, to denote the process opposite to dimensional reduction.
Extended Supersymmetries in $1D$ can be used to constrain, see \cite{{glpr},{top},{fkt}}, possible
higher-dimensional supersymmetric theories (for instance, constraining the number of auxiliary fields
in supergravity theories).
\par
A much more ambitious task, see \cite{{fil},{fl}}, would consist in the reconstruction of a higher-dimensional
theory from its one-dimensional supersymmetric data. \par
An $N$-extended (SuperYang-Mills or supergravity) supersymmetric theory in the ordinary $D=4$ Minkowskian
spacetime produces a $1D$ dimensionally-reduced supersymmetric theory with ${\cal N}=4N$ supercharges. On the other hand,
$N=2$ can be obtained from
the dimensional reduction of $D=6$ (SYM or sugra), $N=4$ from the dimensional reduction of $D=10$ (SYM or sugra) and $N=8$
from the dimensional reduction of the $D=11$ sugra. As a result, a {\em necessary condition} for higher-dimensional oxidation
consists in producing large ${\cal N}$-Extended supersymmetric theories in $1D$. Following \cite{kt2}, the word {\em oxidation}
has been here consistently used in a specific and restricted sense, referring to the operation of enlarging the number of extended
supersymmetries (from ${\cal N}$ to ${\cal N}+1$) acting on a supermultiplet with the same number of component fields.  As such,
the non-minimal
${\cal N}=4$ linear supermultiplets are progressively oxidized to minimal ${\cal N}=5,6,7,8$ linear supermultiplets possessing
$8$ bosonic and $8$ fermionic component fields. It is clear, from these considerations, that the ${\cal N}=4$ off-shell invariant actions based
on non-minimal supermultiplets are not of mere academic interest. Indeed, an $N=2$, $D=4$ theory, dimensionally reduced to $1D$,
produces a supersymmetric model with ${\cal N}=8$ extended supersymmetries; on the other hand the partial spontaneous breaking
of $N=2$ into $N=1$ produces an ${\cal N}=4$ invariant $1D$ supersymmetric model whose component fields belong to ${\cal N}=8$
supermultiplets and are therefore non-minimal supermultiplets w.r.t. the ${\cal N}=4$ invariant supersymmetries.
The inequivalent ${\cal N}=4$ non-minimal supermultiplets and their
inequivalent ${\cal N}=4$-invariant off-shell actions can therefore be regarded as building blocks for constructing
supersymmetric models obtained from dimensional reduction of partial spontaneous supersymmetry breaking of $N=2$, $D=4$
supersymmetry. \par
At the end it is worth mentioning a recent paper \cite{bks} in which the ${\cal N}=4$-invariance for a non-minimal supermultiplet
in presence of a Yang monopole is discussed (see also \cite{gknty}).

{}~
\\{}~
{\Large{\bf Appendix {\bf A}: definitions and conventions}}
~\\
~\par
For completeness we report the definitions, applied to the cases used in the text, of the properties
characterizing the linear representations of the one-dimensional ${\cal N}$-Extended Superalgebra.
In particular the notions of {\em mass-dimension}, {\em field content}, {\em dressing transformation},
{\em connectivity symbol}, {\em dual supermultiplet} and so on, as well as the association of linear
supersymmetry transformations with graphs, will be reviewed following \cite{{pt},{krt},{fg},{kt},{kt2},{grt}}.
The Reader can consult these papers for broader definitions and more detailed discussions. \par
{\em Mass-dimension}:\par
A grading, the mass-dimension $d$, can be assigned to any field entering a linear representation
(the hamiltonian $H$, proportional to the time-derivative operator $\partial\equiv \frac{d}{dt}$,
has a mass-dimension
$1$). Bosonic (fermionic) fields have integer (respectively, half-integer) mass-dimension. \par
{\em Field content:}\par
Each finite
linear representation is characterized by its ``field content", i.e. the set of integers $(n_1,n_2,\ldots , n_l)$
specifying the number $n_i$ of fields of mass-dimension
$d_i$ ($d_i = d_1 + \frac{i-1}{2}$, with $d_1$ an arbitrary constant) entering the representation.
Physically, the $n_l$ fields of highest dimension are the auxiliary fields which transform as a time-derivative
under any supersymmetry generator. The maximal value $l$ (corresponding to the maximal dimensionality
$d_l$) is defined to be the {\em length} of the representation (a root representation has length $l=2$).
Either $n_1, n_3,\ldots$ correspond to the bosonic fields (therefore $n_2, n_4, \ldots$ specify the fermionic
fields)
or viceversa. \\ In both cases the equality $n_1+n_3+\ldots =n_2+n_4+\ldots = n$ is guaranteed.\par
\par
{\em Dressing transformation:}\par
Higher-length supermultiplets are obtained by applying a dressing transformation to the
length-$2$ root supermultiplet. The root supermultiplet is specified by the ${\cal N}$ supersymmetry
operators ${\widehat Q}_i$ ($i=1,\ldots ,{\cal N})$, expressed in matrix form as
\begin{eqnarray}\label{hatq}
{\widehat Q}_j = \frac{1}{\sqrt{2}}\left(
\begin{array}{cc}
0&\gamma_j\\
-\gamma_j\cdot H&0
\end{array}
\right), &&
{\widehat Q}_{\cal N} =\frac{1}{\sqrt{2}} \left(
\begin{array}{cc}
0&{\bf 1}_n\\
{\bf 1}_n\cdot H&0
\end{array}
\right),
\end{eqnarray}
where  the $\gamma_j$ matrices ($j=1,\ldots , {\cal N}-1$) satisfy the Euclidean Clifford algebra
\bea
\{\gamma_i,\gamma_j\}&=& -2\delta_{ij}{\bf 1}_n.
\eea
The length-$3$ supermultiplets are specified by the ${\cal N}$ operators $Q_i$, given by the dressing transformation
\begin{eqnarray}\label{dressingtransformation}
Q_i &=& D{\widehat Q}_i D^{-1},
\end{eqnarray}
where $D$ is a diagonal dressing matrix such that
\bea\label{dressingmatrix}
{D} &=& \left(
\begin{array}{cc}
{\widetilde{D}}&0\\
0&{\bf 1}_n
\end{array}
\right),
\end{eqnarray}
with ${\widetilde D}$ an $n\times n$ diagonal matrix whose diagonal entries are either $1$ or the derivative
operator $\partial$. \par
{\em Association with graphs:}\par
The association between linear supersymmetry transformations and ${\cal N}$-colored oriented graphs goes as
follows. The fields (bosonic and fermionic) entering a representation
are expressed as vertices. They can be accommodated into an $X-Y$ plane. The $Y$ coordinate
can be chosen to
correspond to the mass-dimension $d$ of the fields. Conventionally, the lowest dimensional fields
can be
associated to vertices lying on the $X$ axis. The higher dimensional fields have positive, integer or
half-integer values of $Y$.
A colored edge links two vertices which are connected by a supersymmetry transformation. Each one of the ${\cal N}$ $Q_i$
supersymmetry generators is associated to a given color. The edges are oriented. The orientation reflects the sign
(positive or negative) of the corresponding supersymmetry transformation connecting the two vertices. Instead
of using
arrows, alternatively, solid or dashed lines can be associated, respectively, to positive or negative signs.
No colored line is drawn for supersymmetry transformations connecting a field with the time-derivative of a
lower
dimensional field. This is in particular true for the auxiliary fields (the fields of highest dimension in the
representation) which are necessarily mapped, under supersymmetry transformations, in the time-derivative
of lower-dimensional fields.
\par
Each irreducible supersymmetry transformation can be presented (the identification is not unique) through
an oriented
${\cal N}$-colored graph with $2n$ vertices. The graph is such that precisely ${\cal N}$ edges, one for each
color, are linked to any given vertex which represents either a $0$-mass dimension or a $\frac{1}{2}$-mass
dimension field.  An unoriented ``color-blind" graph can be associated to the initial graph by disregarding
the orientation of the edges
and their colors (all edges are painted in black). \par
{\em Connectivity symbol:}\par
A characterization of length $l=3$ color-blind, unoriented graphs can be expressed through the connectivity
symbol
$\psi_g$, defined as follows
\begin{eqnarray}
\psi_g &=& ({m_1})_{s_1} +({m_2})_{s_2}+\ldots +({m_Z})_{s_Z}.
\end{eqnarray}
The $\psi_g$ symbol encodes the information on the partition of the $n$  $\frac{1}{2}$-mass dimension fields
(vertices)
into the sets of $m_z$ vertices ($z=1,\ldots, Z$) with $s_z$ edges connecting them to the $n-k$ $1$-mass
dimension auxiliary fields.
We have
\begin{eqnarray}
m_1+m_2+\ldots +m_Z &=& n,
\end{eqnarray}
while $
s_z\neq s_{z'}$ for $  z\neq z'$.\par
{\em Dual supermultiplet:}\par
A dual supermultiplet is obtained by mirror-reversing, upside-down, the graph associated
to the original supermultiplet.

{}~
\\{}~
{\Large{\bf Appendix {\bf B}: explicit construction of non-minimal supermultiplets}}
~\\
~\par
We present here an explicit construction of the non-minimal, reducible but indecomposable ${\cal N}=4$
supermultiplets of length $l=2,3$.\par
We start with the ${\cal N}=8$ $(8,8)$ root supermultiplet expressed through (\ref{hatq}), with the $7$
matrices
$\gamma_j$ given by
\begin{center}
\begin{tabular}{llll}
$\gamma_1 = \tau_1\otimes{\bf 1}_2\otimes\tau_A, $&$\gamma_2 = \tau_2\otimes{\bf 1}_2\otimes\tau_A, $&$
\gamma_3 = \tau_A\otimes \tau_1\otimes{\bf 1}_2,$&$\gamma_4 = \tau_A\otimes \tau_2\otimes{\bf 1}_2, $\\
$\gamma_5 = {\bf 1}_2\otimes\tau_A\otimes\tau_1, $&$\gamma_6 = {\bf 1}_2\otimes\tau_A\otimes\tau_2, $&$
\gamma_7 = \tau_A\otimes\tau_A\otimes\tau_A,$&$ $
\end{tabular}
\end{center}
where $\tau_1$, $\tau_2$, $\tau_A$, ${\bf 1}_2$ are $2\times 2$ matrices given by
($e_{mn}$ is the $2\times 2$ matrix with entry $1$ at the $m^{th}$ row, $n^{th}$ column
and $0$ otherwise)
\bea
&\tau_1 = e_{12}+e_{21}, \quad \tau_2 = e_{11}-e_{22},\quad \tau_A = e_{12}-e_{21},\quad{\bf 1}_2
=
e_{11}+e_{22}. \nonumber
\eea
We can select, e.g., the $4$ operators producing the non-minimal ${\cal N}=4$ supermultiplet of length $l=2$
(with connected graph) to be given by ${\widehat Q_2}$, ${\widehat Q_5}$, ${\widehat Q_6}$, ${\widehat Q_7}$.
For this choice of root operators, the dressing transformations (\ref{dressingtransformation}) producing the
inequivalent length $l=3$ non-minimal,
reducible but indecomposable ${\cal N}=4$ supermultiplets are obtained by applying the following diagonal
dressing matrices ${\widetilde D}$, see (\ref{dressingmatrix}):
\begin{center}
\begin{tabular}{|c|c|c|}\hline
\emph{field content:}& \emph{label:}& \emph{dressing matrix ${\widetilde D}$:}\\ \hline
$(1,8,7)$ &&$diag(1,\partial,\partial,\partial,\partial,\partial,\partial,\partial)$  \\ \hline
$(2, 8, 6)$& $A$& $diag(1,1,\partial,\partial,\partial,\partial,\partial,\partial)$\\ \cline{2-3}
& $B$& $diag(1,\partial,\partial,\partial,\partial,\partial,\partial,1)$\\ \hline
$(3, 8, 5)$ &$A$& $diag(1,1,1,\partial,\partial,\partial,\partial,\partial)$\\ \cline {2-3}
&$B$& $diag(1,1,\partial,\partial,\partial,\partial,\partial,1)$\\ \hline
$(4, 8, 4)$& $A$& $diag(\partial,1,1,1,1,\partial,\partial,\partial)$\\ \cline {2-3}
&$B$& $diag(1,1,1,1,\partial,\partial,\partial,\partial)$\\ \cline {2-3}
&$C$& $diag(1,1,1,\partial,1,\partial,\partial,\partial)$\\ \cline {2-3}
&$D$& $diag(1,1,\partial,\partial,\partial,\partial,1,1)$\\ \hline
$(5, 8, 3)$& $A$& $diag(1,1,1,1,1,\partial,\partial,\partial)$\\ \cline{2-3}
& $B$& $diag(\partial,1,1,1,1,1,\partial,\partial)$\\ \hline
$(6, 8, 2)$& $A$& $diag(1,1,1,1,1,1,\partial,\partial)$\\ \cline{2-3}
& $B$& $diag(\partial,1,1,1,1,1,1,\partial)$\\ \hline
$(7, 8, 1)$ &&$diag(1,1,1,1,1,1,1,\partial)$\\ \hline
\end{tabular}
\end{center}
\bea&&\eea

\newpage
{}~
\\{}~
\par {\large{\bf Acknowledgments}}{} ~\\{}~\par
M. G. receives a CNPq/CLAF grant. S. K. receives a CAPES grant.
The work has been supported by Edital Universal CNPq, Proc. 472903/2008-0.

\

{}~
\\{}~
\newpage
{\Large{\bf Tables with a few selected (unoriented, color-blind) graphs}}
~\\
~\par
For sake of clarity we present a few selected (unoriented, color-blind) graphs associated
to the non-minimal,
both reducible but indecomposable and fully reducible, linear supermultiplets.\par
We list the graphs associated to the inequivalent ${\cal N}=4$ non-minimal linear supermultiplets
of field content $(8,8)$ and $(2,8,6)$, respectively. The graphs associated to the length-$3$ supermultiplets encode,
in particular, the information of their connectivity symbol.
We have
\newpage
\begin{figure}
  \centering
  \includegraphics[width=0.9\textwidth]{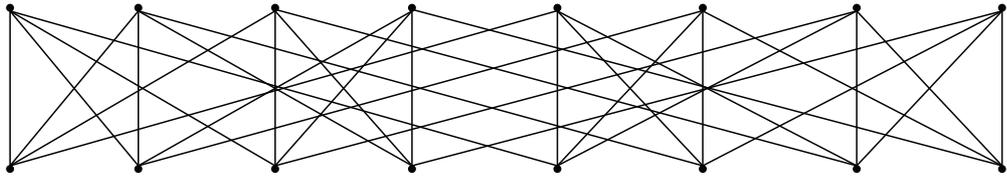}\\
   \caption{$(8,8)_{red}$, reducible but indecomposable.}
\end{figure}
\begin{figure}
   \centering
  \includegraphics[width=0.9\textwidth]{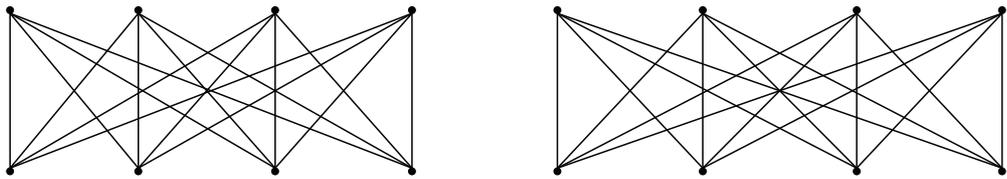}\\
   \caption{$(8,8)_{FR}$, fully reducible.}
\end{figure}
\begin{figure}
  \centering
  \includegraphics[width=0.9\textwidth]{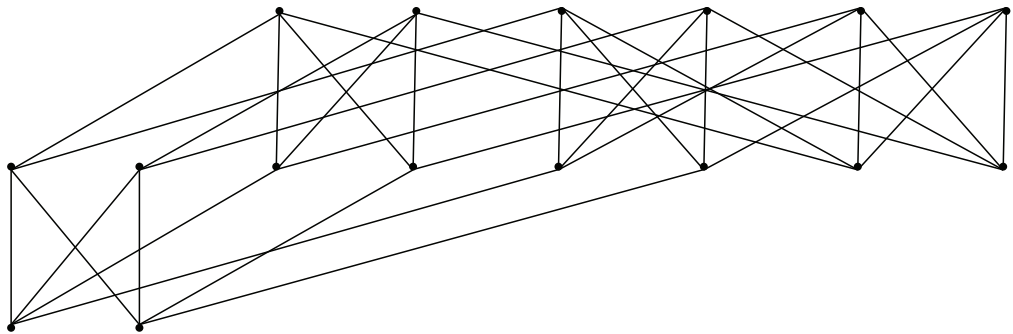}\\
   \caption{$(2,8,6)_A$, reducible but indecomposable (connectivity symbol $2_4+4_3+2_2$).}
\end{figure}
\begin{figure}
  \centering
  \includegraphics[width=0.9\textwidth]{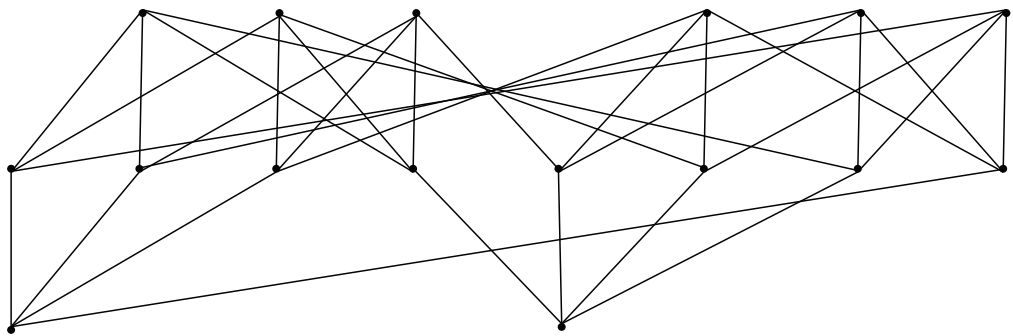}\\
   \caption{$(2,8,6)_B$, reducible but indecomposable (connectivity symbol $8_3$).}
\end{figure}
\begin{figure}
  \centering
  \includegraphics[width=0.9\textwidth]{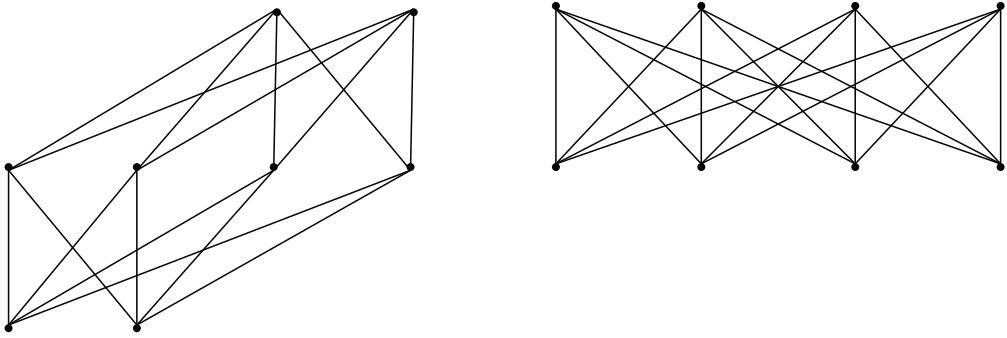}\\
   \caption{$(2,8,6)_a$, fully reducible (connectivity symbol $4_4+4_2$).}
\end{figure}
\begin{figure}
  \centering
  \includegraphics[width=0.9\textwidth]{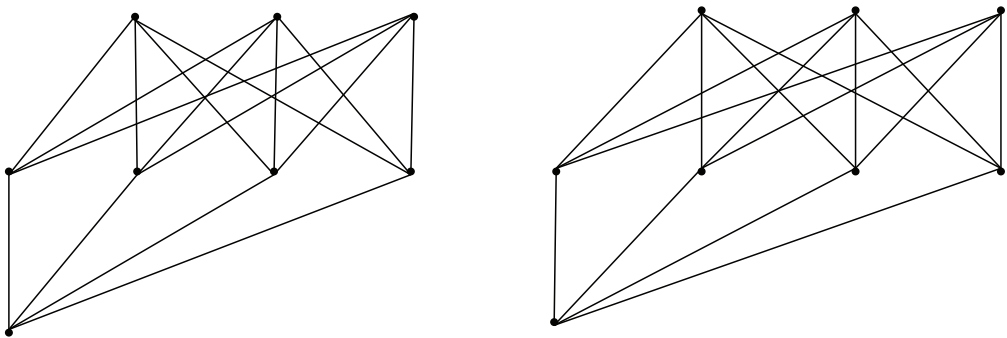}\\
   \caption{$(2,8,6)_b$, fully reducible (connectivity symbol $8_3$).}
\end{figure}

\end{document}